\pdfoutput=1
\documentclass[
 aip,
 pop,
 amsmath,
 amssymb,
 reprint,
 floatfix,
 raggedbottom
]{revtex4-2}

\usepackage{amsmath, amssymb, bm}
\usepackage{graphicx}
\usepackage{booktabs}
\usepackage{siunitx}
\usepackage{multirow}
\usepackage{placeins}
\usepackage{tabularx,array,float}
\usepackage{hyperref}
\usepackage{xcolor}

\newcommand{\figwideone}{6.69in}

\hypersetup{
    colorlinks=true,
    linkcolor=blue,
    citecolor=blue,
    urlcolor=blue
}

\AtBeginDocument{%
}

\providecommand{\FreezeFZeroRecordCount}{197}
\providecommand{\FreezeFZeroRecordMeanKHz}{39.360}
\providecommand{\FreezeFZeroRecordMedianKHz}{39.300}
\providecommand{\FreezeFZeroRecordMinKHz}{38.570}
\providecommand{\FreezeFZeroRecordMaxKHz}{40.820}
\providecommand{\FreezeTwoFZeroRecordMinKHz}{77.14}
\providecommand{\FreezeTwoFZeroRecordMaxKHz}{81.64}
\providecommand{\FreezeTargetBinCount}{4}
\providecommand{\FreezeResidualNullCount}{195}
\providecommand{\FreezeResidualNullTotal}{197}
\providecommand{\FreezeBicoXRecordMedian}{0.9624}
\providecommand{\FreezeBicoCRecordMedian}{0.9991}
\providecommand{\FreezeBicoRRecordMedian}{0.02654}
\providecommand{\FreezePhasePowerClosureMax}{8.66\times10^{-15}}

\providecommand{\FreezeSignedPrimaryTotal}{0.02234582}
\providecommand{\FreezeSignedStrictTotal}{0.01795684}
\providecommand{\FreezeSignedPrimaryCCCFrac}{0.91494}
\providecommand{\FreezeSignedStrictCCCFrac}{0.95145}
\providecommand{\FreezeSignedPrimaryCCCFracMinimum}{0.87311}
\providecommand{\FreezeSignedStrictCCCFracMinimum}{0.90871}
\providecommand{\FreezeJackknifePrimaryLowerMinimum}{0.84857}
\providecommand{\FreezeJackknifeStrictLowerMinimum}{0.85675}
\providecommand{\FreezeSignedPrimaryPositiveCount}{20}
\providecommand{\FreezeSignedStrictPositiveCount}{20}
\providecommand{\FreezeSignedPrimaryCCCDominanceCount}{20}
\providecommand{\FreezeSignedStrictCCCDominanceCount}{20}
\providecommand{\FreezeSignedPrimaryResidualPositiveCount}{13}
\providecommand{\FreezeSignedPrimaryOneR}{1.24\times10^{-4}}
\providecommand{\FreezeSignedPrimaryTwoR}{2.40\times10^{-3}}
\providecommand{\FreezeSignedPrimaryRRR}{-5.64\times10^{-4}}
\providecommand{\FreezeExcludeTwoFCCCMedian}{0.7945}
\providecommand{\FreezeExcludeTwoFCCCMinimum}{0.6910}
\providecommand{\FreezeShiftFortyEightyCCCMedian}{0.9410}
\providecommand{\FreezeShiftFortyEightyCCCMinimum}{0.9000}
\providecommand{\FreezeShiftFiftyNinetyCCCMedian}{0.9992}
\providecommand{\FreezeShiftFiftyNinetyCCCMinimum}{0.9670}
\providecommand{\FreezeHigherTwoThreeCount}{20}
\providecommand{\FreezeHigherThreeFourCount}{19}
\providecommand{\FreezeEnvelopeNullCount}{19}
\providecommand{\FreezeEnvelopeNullTotal}{20}
\providecommand{\FreezeMZeroRealizations}{50}
\providecommand{\FreezeMOneRealizations}{200}
\providecommand{\FreezeMZeroBandCoverage}{59/300}
\providecommand{\FreezeMOneBandCoverage}{66/300}
\providecommand{\FreezeMZeroNotZeroCoverage}{0/60}
\providecommand{\FreezeMOneNotZeroCoverage}{0/60}
\providecommand{\FreezeMZeroZeroOddCoverage}{13/100}
\providecommand{\FreezeMOneZeroOddCoverage}{13/100}
\providecommand{\FreezeMZeroCCCCoverage}{0/40}
\providecommand{\FreezeMOneCCCCoverage}{0/40}

\providecommand{\FreezeMOnePeakObservedByBand}{1.143, 1.205, 2.153, 1.792, 1.301}
\providecommand{\FreezeMOnePeakGeneratedByBand}{1.046, 1.099, 1.837, 1.838, 1.165}
\providecommand{\FreezeMZeroSubMHzPeakGenerated}{1.061, 1.062}
\providecommand{\FreezeMOnePsdXMedian}{4.19\times10^{-7}}
\providecommand{\FreezeMOnePsdYMedian}{4.39\times10^{-7}}
\providecommand{\FreezeMOneCrossMedian}{4.27\times10^{-7}}
\providecommand{\FreezeMOneCovarianceMaximum}{1.03\times10^{-6}}
\providecommand{\FreezeBicoXObserved}{0.9631}
\providecommand{\FreezeBicoXMOne}{0.9233}
\providecommand{\FreezeBicoRObserved}{0.02723}
\providecommand{\FreezeBicoRMOne}{0.000698}
\providecommand{\FreezeBicoRObservedOverMOne}{39.0}

\providecommand{\FreezeCCCObserved}{0.02144}
\providecommand{\FreezeCCCMOne}{0.01406}
\providecommand{\FreezeOneRObserved}{1.24\times10^{-4}}
\providecommand{\FreezeOneRMOne}{-1.56\times10^{-6}}
\providecommand{\FreezeTwoRObserved}{0.002401}
\providecommand{\FreezeTwoRMOne}{-4.30\times10^{-6}}
\providecommand{\FreezeRRRObserved}{-5.64\times10^{-4}}
\providecommand{\FreezeRRRMOne}{-1.81\times10^{-8}}

\providecommand{\FreezePowerTableRows}{%
60--100~kHz & 0.8794 & 0.1246 & -0.0046 & 1.168 & 0/20 \\
100--200~kHz & 0.5538 & 0.4496 & -0.0010 & 1.235 & 18/20 \\
0.2--0.5~MHz & 0.1428 & 0.8714 & -0.0185 & 2.370 & 20/20 \\
0.5--1~MHz & 0.0020 & 0.9974 & 0.000367 & 1.981 & 20/20 \\
1--3~MHz & 0.0000768 & 0.99995 & -0.0000305 & 1.164 & 19/20 \\
}

\providecommand{\SIInstrumentRangeRows}{%
Fixed probe (C1) & 197 & 11,664 & 7,007 & 0 & 0 \\
Movable probe (C3) & 197 & 14,224 & 9,599 & 0 & 0 \\
}

\providecommand{\SIAcquisitionOrderRows}{%
Breathing frequency, $f_0$ (kHz) & -0.042718 & [-0.10549, 0.017728] & No \\
Residual fraction, 60--100 kHz & $9.38\times10^{-4}$ & [-0.003656, 0.0068022] & No \\
Residual fraction, 100--200 kHz & -0.001553 & [-0.00634, 0.0032795] & No \\
Residual fraction, 0.2--0.5 MHz & 0.0025432 & [-0.0048436, 0.012852] & No \\
Residual fraction, 0.5--1 MHz & $-3.9\times10^{-4}$ & [-0.0010409, $2.66\times10^{-4}$] & No \\
Residual fraction, 1--3 MHz & $-2.15\times10^{-5}$ & [$-5.57\times10^{-5}$, $1.37\times10^{-5}$] & No \\
Residual bicoherence, $b_R^2$ & 0.02223 & [-0.0069221, 0.064684] & No \\
$S_{CCC}/S_{\mathrm{total}}$ & 0.0051593 & [-0.011563, 0.026577] & No \\
}

\newcommand{\alttext}[1]{\par\smallskip{\small\raggedright\textbf{Alternative text:} #1\par}}

\begin{document}

\title{Contributions of the breathing waveform and residual fluctuations to power and higher-order spectra in a Hall-thruster plume}

\author{Zilong Peng}
\affiliation{School of Energy Science and Engineering, Harbin Institute of Technology, Harbin 150001, People's Republic of China}
\author{Baisheng Wang}
\affiliation{School of Energy Science and Engineering, Harbin Institute of Technology, Harbin 150001, People's Republic of China}
\author{Zhongping Zhao}
\affiliation{School of Energy Science and Engineering, Harbin Institute of Technology, Harbin 150001, People's Republic of China}
\author{Yao Tang}
\affiliation{School of Energy Science and Engineering, Harbin Institute of Technology, Harbin 150001, People's Republic of China}
\author{Yinjian Zhao}
\email{zhaoyinjian@hit.edu.cn}
\affiliation{School of Energy Science and Engineering, Harbin Institute of Technology, Harbin 150001, People's Republic of China}
\author{Daren Yu}
\affiliation{School of Energy Science and Engineering, Harbin Institute of Technology, Harbin 150001, People's Republic of China}
\author{Ming Zeng}
\affiliation{School of Energy Science and Engineering, Harbin Institute of Technology, Harbin 150001, People's Republic of China}
\date{October 6, 2026}

\begin{abstract}
Higher-frequency power in Hall-thruster plume-probe signals can arise from
both harmonics of a repeatable breathing waveform and fluctuations about it,
complicating spectral interpretation. We decompose 197 paired records from
20 sequentially sampled probe-pair geometries at one operating condition,
estimating each phase-conditioned waveform from other records at the same
geometry. The residual accounts for 87.1\% and 99.7\% of measured power in
the 0.2--0.5 and 0.5--1~MHz bands, respectively, as medians across geometries.
The repeatable waveform, however, reproduces the principal
breathing-harmonic bicoherence structure. Residual power remains modulated
by breathing phase after waveform removal, exceeding geometry-specific
95th-percentile circular-shift thresholds in both bands at all 20 geometries.
Residual bicoherence is substantially weaker than that of the measured
signal but exceeds record-specific random-phase thresholds in 195 of 197
records. An exact component expansion further shows that the all-waveform
term dominates a specified signed third-order statistic over the fixed
50--80~kHz band near the second breathing harmonic. These results distinguish
band-power allocation from harmonic organization within the evaluated
frequency ranges. The attribution concerns uncalibrated probe-chain signals
and does not identify distinct physical modes or measure energy-transfer
rates. The findings motivate assessing repeatable-waveform contributions
before interpreting breathing-harmonic bicoherence as evidence of
interactions among residual fluctuations.
\end{abstract}


\maketitle

% ============================================================
\section{Introduction}
\label{sec:introduction}
% ============================================================

Hall-thruster plasmas exhibit low-frequency breathing oscillations together
with fluctuations at higher frequencies. The breathing oscillation is
associated with the coupled depletion and replenishment of neutral
propellant, providing a periodic background for the measured fluctuations
\cite{boeuf2017,choueiri2001,kaganovich2020}. Time-resolved measurements of
ion, neutral, and electron properties \cite{dale2021}, including probe
measurements of density, potential, and electron temperature
\cite{giannetti2020} and phase-resolved ion velocity distributions
\cite{dancheva2023}, establish breathing phase as a useful coordinate for
examining plasma dynamics within the oscillation cycle.

Faster fluctuations can also vary across that cycle: high-frequency bursts
in the cathode potential have been observed at particular stages of the
discharge-current oscillation \cite{albarede2006}. Spectral measurements
further show that breathing harmonics can overlap the frequency range
occupied by faster fluctuations \cite{delaviere2024}. Higher-frequency power
can therefore include both harmonics of the repeatable breathing waveform
and fluctuations about that waveform. Distinguishing these contributions is
necessary before assessing how fluctuations beyond the repeatable waveform
are organized over the breathing cycle. Phase dependence alone cannot make
this distinction, because both the waveform and the intensity of residual
fluctuations can vary with breathing phase.

A repeatable, non-sinusoidal waveform contains phase-locked harmonics that
can produce strong bispectral structure without requiring interactions among
distinct stochastic modes \cite{kovach2018,lozano2016,giehl2021}.
Kobayashi et al.\ observed significant fundamental--harmonic bicoherence
alongside a non-sinusoidal temperature-fluctuation waveform in a toroidal
plasma \cite{kobayashi2012}. Bicoherence thus quantifies cross-frequency
phase organization, while identifying a generation mechanism or a direction
of energy transfer requires additional evidence
\cite{kimpowers1979,itoh2005,wang2024}. Spatially resolved approaches provide
complementary information through measured azimuthal mode numbers
\cite{park2025}, frequency--wavenumber-resolved bicoherence in simulations
\cite{bayon2025}, and two-point bispectral analyses of spoke harmonics and
wave decay \cite{przybocki2025}. For interpreting breathing-harmonic
bicoherence, a further question is how much of the measured signature can
be reproduced by the repeatable waveform itself.

Separating a periodic mean from residual fluctuations has an established
basis in cyclostationary analysis \cite{antoni2009}. Raad et al.\ emphasize
that higher-order moments can contain contributions inherited from
lower-order moments, and that the relevant mean varies over the cycle
\cite{raad2008}. Phase-resolved observations alone, however, do not quantify
how the repeatable breathing waveform and its residual contribute differently
to spectral power and higher-order statistics in the same measured records.

Here, we quantify waveform and residual contributions to higher-frequency
band power in paired Hall-thruster plume-probe records, then examine the
breathing-phase dependence of residual power. We compare these results with
the waveform and residual bicoherence to distinguish power allocation from
harmonic phase organization. A leave-one-record-out phase-conditioned
decomposition constructs the waveform from other records at the same probe
geometry, supplying common components for these comparisons. An exact
trilinear expansion further attributes a specified signed third-order
statistic over a fixed frequency band near the second breathing harmonic.
These statistical attributions do not identify distinct physical modes or
measure an energy-transfer rate.

Section~\ref{sec:experiment} describes the experimental setup and data
acquisition. Section~\ref{sec:methodology} presents the decomposition and
statistical comparisons. The results are reported in Sec.~\ref{sec:results},
followed by the discussion and conclusions in Secs.~\ref{sec:discussion}
and~\ref{sec:conclusions}.

% ============================================================
\section{Experimental Setup and Data Acquisition}
\label{sec:experiment}
% ============================================================

\begin{figure*}[!tp]
  \centering
  \includegraphics[width=\figwideone]{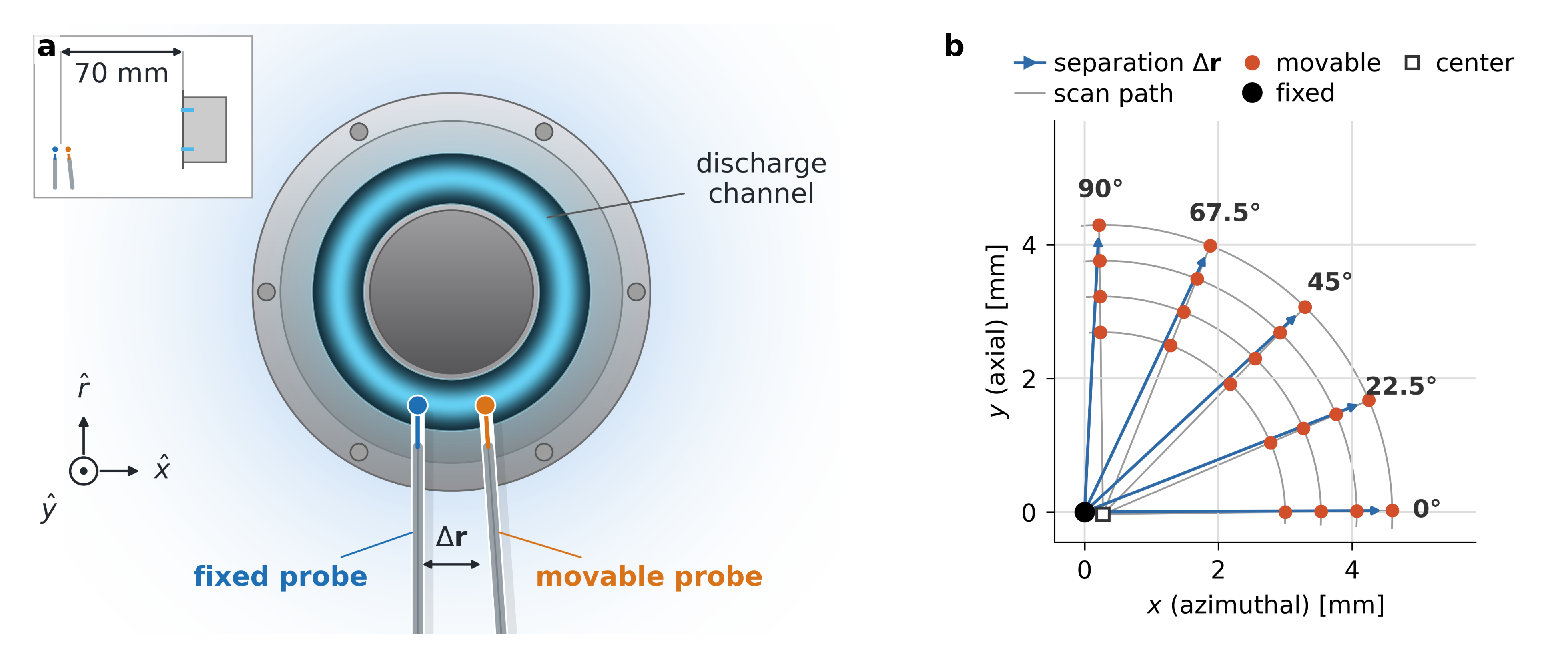}
  \caption{Measurement station and reconstructed probe-pair geometry.
  (a) Upstream-looking projection of the Hall-thruster exit plane; the probe
  spacing is enlarged relative to the channel width. The inset locates the
  probe station 70~mm downstream of the channel exit. The unit vectors
  $\hat{x}$, $\hat{r}$, and $\hat{y}$ denote the azimuthal, radial, and axial
  directions, respectively; $\hat{y}$ points out of the page. (b) Reconstructed
  fixed-to-movable separation vectors for the 20 sequentially acquired
  geometries. The open square marks the stage center, which is offset
  from the fixed probe by 0.28~mm. The reconstructed directions therefore
  differ from the nominal scan-angle labels.}
  \label{fig:experiment-geometry}
\end{figure*}

The probe separation and orientation were varied in preparation for
dispersion measurements using the spatial anti-aliasing approach of Liu and
Jorns.\cite{liu2025antialiasing} The present study uses these paired
measurements to determine how the repeatable breathing waveform and residual
fluctuations contribute to spectral power and higher-order statistics.
Figure~\ref{fig:experiment-geometry} shows the measurement station and
reconstructed probe-pair geometries.

\subsection{Discharge conditions and measurement station}

The experiment was performed in the plume of a permanent-magnet SNEP30 Hall
thruster operated on xenon. The discharge
voltage was 250~V and the instrument-reported anode current was 0.768~A,
corresponding to a discharge power of approximately 200~W. The anode and
cathode xenon flow rates were 8 and 1~sccm, respectively. The facility
background pressure during the measurements was approximately
$1\times10^{-3}$~Pa. The discharge channel was 21.6~mm long, comprising a
5.9~mm gas distributor and a 15.7~mm net discharge length.

The probe station was 70~mm downstream of the channel exit, aligned with the
radial centerline of the annular discharge channel. A Finite Element Method
Magnetics (FEMM) calculation\cite{meeker2015femm} predicted a magnetic-field magnitude of
approximately 9.59~mT at this location. This simulated field value was not
used quantitatively in the present analysis.

\subsection{Probe geometry and data acquisition}
\label{sec:word-conversion}

The two electrostatic probes used 0.5~mm-diameter tungsten wires with
2.5~mm exposed collecting lengths. The fixed probe was vertical and the movable
probe was slightly inclined. Their tips shared the same radial height, placing
the separation vector $\Delta\bm{r}$ in the axial--azimuthal plane.

The scan used five nominal arm rotations: $0^{\circ}$, $22.5^{\circ}$,
$45^{\circ}$, $67.5^{\circ}$, and $90^{\circ}$. Four nominal translation
commands of 0, 0.5, 1.0, and 1.5~mm gave 20 probe-pair geometries in
combination with these rotations. The nominal $0^{\circ}$ direction was
referenced to the local azimuthal direction $x$, and the nominal $90^{\circ}$
direction to the axial direction $y$. The four translation blocks were
acquired sequentially, with the rotation order reversed between successive
blocks rather than randomized.

\begin{figure}[!tbp]
  \centering
  \includegraphics[width=3.37in]{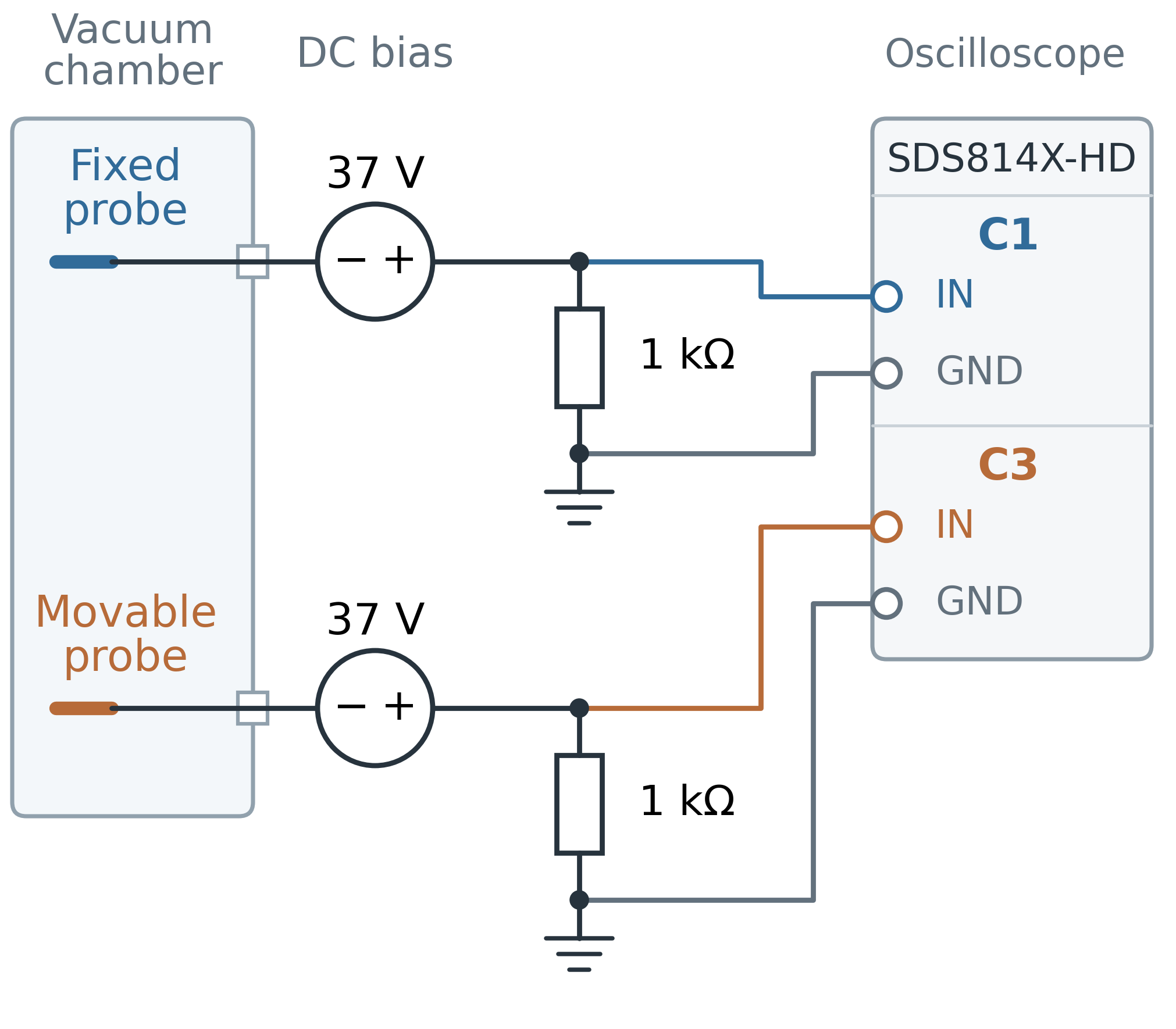}
  \caption{Probe-bias and voltage-drop sensing circuit. The two branches
  show the fixed- and movable-probe sensing connections. The source symbols
  indicate the bias polarity, with negative terminals toward the probes.
  Each channel measures across its sensing resistor, with the input lead
  (IN) connected to the bias-side terminal and the ground lead (GND) to the
  grounded terminal.}
  \label{fig:current-sensing}
\end{figure}

Both probes were negatively biased using an HSPY-300-01 DC power supply
set to $-37$~V and operated nominally in the ion-saturation regime.
Probe-current fluctuations were represented by the voltage drops across
nominal 1~k$\Omega$ sensing resistors. A similar two-probe
voltage-drop sensing topology was described by Liu and
Jorns.\cite{liu2024two_probe} The present bias and acquisition topology is
summarized in Fig.~\ref{fig:current-sensing}.

The fixed- and movable-probe voltage signals were recorded on oscilloscope
channels C1 and C3, respectively. The fixed-probe signal provided the primary breathing-phase
reference for both signals. The 12-bit SDS814X-HD digitized both signals
simultaneously at 100~MS/s for 100~ms per record, giving $10^7$ samples per
channel. The recorded voltage signals were used in the subsequent spectral
analyses.

The acquisition schedule contained ten paired records for each of the 20
geometries. Three of the 200 scheduled pairs were not stored on
either channel, leaving 197 complete paired records and 8--10 records per
geometry. Beyond this completeness requirement, no additional phase-quality
exclusion was applied.

Stored waveform codes were converted to digitizer-input voltage using the
channel-specific calibration parameters in each file header. Independent
decoding and digital-range audits covered all 394 record--channel waveforms
and found no stored-word clipping; the conversion formula and complete audit
results are provided in the supplementary material, Sec.~S1.

No single-probe current--voltage trace or electron temperature was measured
under these operating conditions. The recorded voltages are
therefore treated as relative ion-current fluctuation proxies rather than
absolute plasma-density measurements. The oscilloscope input-impedance
setting, intermediate analog-chain components, analog bandwidth, and trigger
setting were undocumented, and a calibrated sensing resistance for conversion
to absolute current was unavailable. No post-acquisition amplitude- or
phase-response correction was applied to either signal. The analysis therefore
concerns the recorded probe-chain voltage signals, without calibration to
current, density, or a physical transfer rate.

% Start the calibration subsection beside the circuit float in the clean copy.
\newpage % clean-copy calibration column break
\begin{samepage}
\subsection{Geometric calibration and measurement scope}

The nominal angles and translation commands identify the scan settings,
rather than the reconstructed probe separations. Measured zero-translation
tip spacings and an arm-extension calibration defined the reconstructed
separation vectors in Fig.~\ref{fig:experiment-geometry}.
Their lengths ranged from 2.7 to 4.6~mm across the 20 geometries.
The calibration inputs, reconstruction branch, and unavailable mechanical
uncertainties are documented in the supplementary material, Sec.~S1.
\par
\end{samepage}

Illustrative waveforms and spectral maps use the probe-pair configuration
with a nominal scan angle of $45^{\circ}$ and a reconstructed separation of
3.43~mm. The configuration and example record were selected without reference
to their spectral results; the selection rule is given in the supplementary
material, Sec.~S4.4. Statistical summaries use all 197 paired records across
the 20 configurations.

The analysis does not use $\Delta\bm{r}$ as a propagation factor
or perform a wavenumber inversion. Agreement across geometries therefore
measures reproducibility across sequential acquisitions, not a simultaneous
spatial snapshot. The file identifier remains only a within-geometry
acquisition-order proxy because absolute acquisition timestamps were not retained
for this data set.

\begin{figure*}[!tp]
  \centering
  \includegraphics[width=\figwideone]{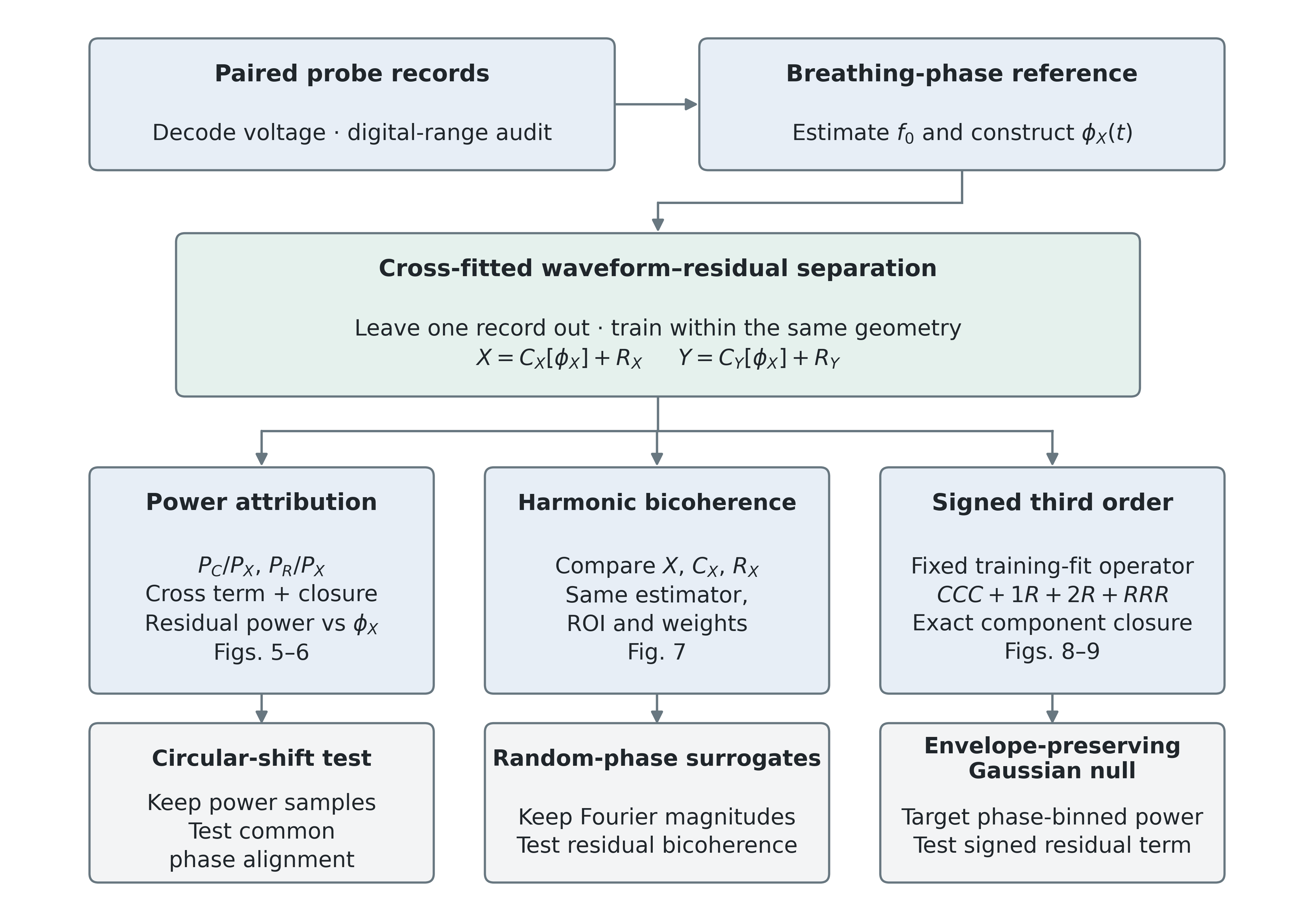}
  \caption{Analysis workflow. A shared cross-fitted decomposition supplies
  three parallel analyses, each with its corresponding statistical comparison.
  Arrows indicate analysis dependence rather than physical causality;
  robustness checks are detailed in the supplementary material, S3--S7.}
  \label{fig:analysis-pipeline}
\end{figure*}

\FloatBarrier

% ============================================================
\section{Analysis Methodology}
\label{sec:methodology}
% ============================================================

This section presents the framework for attributing spectral power and
higher-order statistics to the repeatable breathing waveform and residual
fluctuations. It defines the signal decomposition and the estimators used for
power accounting, bicoherence comparison, and signed third-order attribution.
The accompanying statistic-specific surrogate tests assess the statistical
structure retained in the residual.

\begin{figure*}[!tp]
  \centering
  \includegraphics[width=\figwideone]{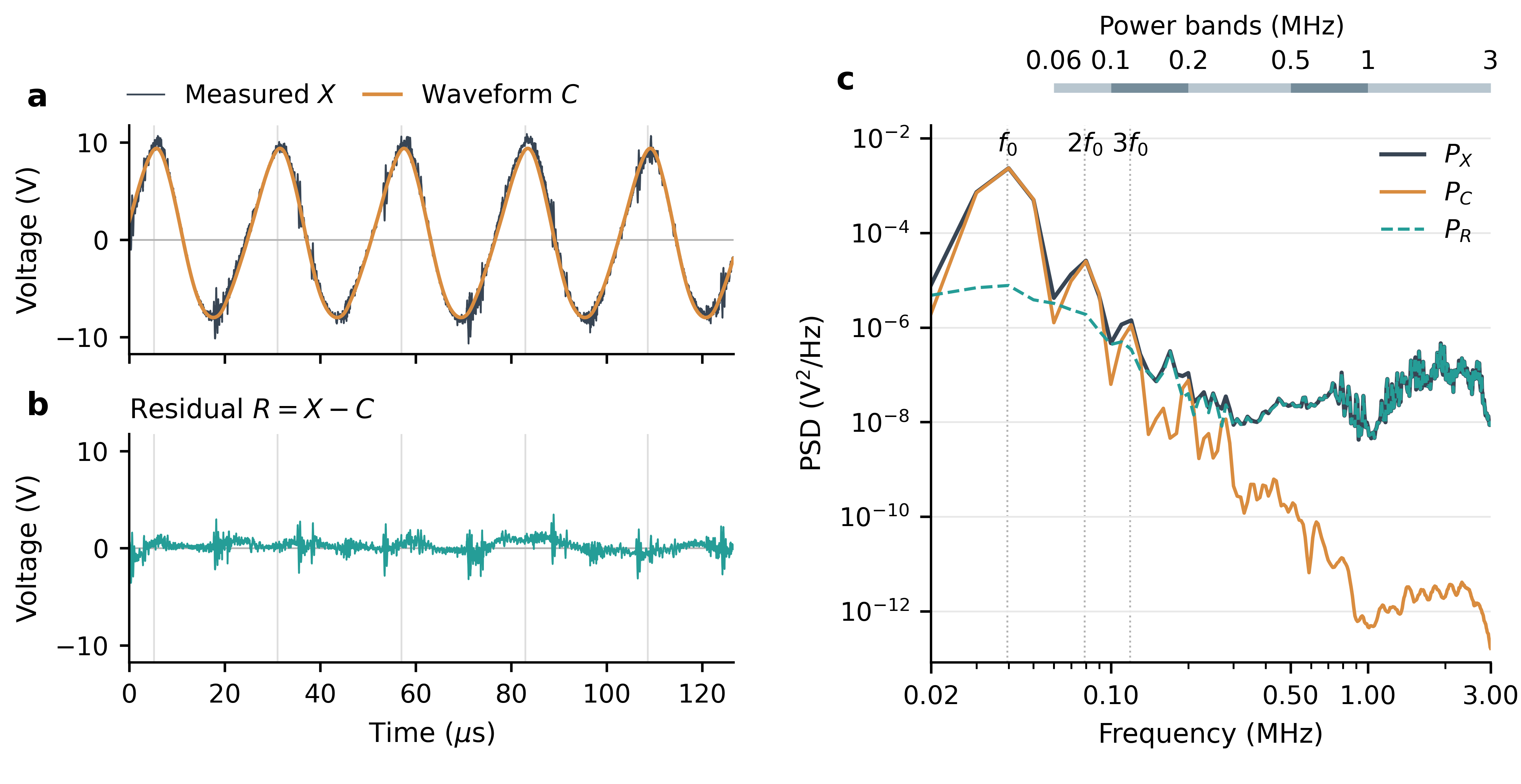}
  \caption{Measured signal and cross-fitted waveform--residual decomposition
  for one held-out fixed-probe record at a nominal scan angle of
  $45^{\circ}$ and a reconstructed probe separation of 3.43~mm.
  (a) Measured signal and waveform; (b) residual, on identical voltage and
  time scales. Here $C=C_X$ and $R=R_X$.
  The window starts at 50~ms and spans five breathing periods
  (126.55~$\mu$s); vertical guides mark integer cycles.
  Traces are full-record demeaned, without smoothing or amplitude rescaling.
  (c) One-sided power spectral densities (PSD) from the full
  100~ms record, using 1,000 non-overlapping, mean-removed 10,000-point
  Hann-windowed segments. The dotted guides mark the first three harmonics
  of $f_0=39.51$~kHz; the top strip identifies the five power-analysis bands.}
  \label{fig:waveform-residual-example}
\end{figure*}

\subsection{Signal definitions and analysis overview}

Let $X(t)$ and $Y(t)$ denote the fixed- and movable-probe signals,
respectively, recorded as described in Sec.~\ref{sec:experiment}.
Their Fourier coefficients are written as $X(f)$ and $Y(f)$. The breathing
phase $\phi_X(t)$ extracted from the fixed-probe signal provides the common
reference for decomposing both signals.

A complete paired record is the primary analysis unit. The 20 sequential
probe-pair geometries define the groups for summaries and cluster resampling. These
summaries assess reproducibility across scan settings; the geometries do not
constitute independent repetitions of the experiment. Breathing cycles,
spectral segments, and frequency pixels are used within records and are not
treated as independent experimental samples.

Figure~\ref{fig:analysis-pipeline} summarizes three complementary analyses of
the same waveform and residual components. Band-power analysis determines
which component carries the power and whether residual power varies with
breathing phase. Bicoherence comparisons assess which component reproduces
the measured harmonic organization. The signed third-order analysis provides
an exact component expansion of a specified statistic. Each analysis uses
the corresponding statistic-specific comparison shown in the figure.
The supplementary conditional comparison with Gaussian controls is reported
separately in supplementary material, Sec.~S6.

\subsection{Breathing phase and cross-fitted decomposition}

For each evaluated record, we used the other complete records acquired at the
same probe geometry to estimate the signal that typically recurred at each
breathing phase. Evaluating this phase-conditioned mean along the held-out
record's phase trajectory defines its repeatable breathing waveform;
subtracting that waveform defines the residual. For held-out record $r$ in
geometry $g$, the channel-specific decomposition is
\begin{align}
 X_{g,r}(t)&=C_{X,g,-r}[\phi_{X,g,r}(t)]+R_{X,g,r}(t),\nonumber\\
 Y_{g,r}(t)&=C_{Y,g,-r}[\phi_{X,g,r}(t)]+R_{Y,g,r}(t).
 \label{eq:crossfit-decomposition}
\end{align}
Thus $C_X$ and $C_Y$ denote the waveforms, whereas $R_X$ and $R_Y$ are their
corresponding residuals. Record $r$ contributes no samples to the
phase-binned training means. It still supplies its own phase trajectory and
the breathing-frequency estimate used to select the retained template
harmonics, as specified below.

A common phase convention is required to compare waveform content at the same
position in the breathing cycle across records. The primary phase coordinate
$\phi_X(t)$ was extracted separately for each record from the fixed-probe
signal and used to condition both probe signals. The complete demeaned record
was Fourier transformed. The maximum-power discrete Fourier bin between
35 and 45~kHz defined the record-level breathing-frequency
estimate $f_0$. The Fourier coefficients in the rectangular 25--55~kHz
positive-frequency band were inverse transformed to a complex baseband
signal. After removal of the baseband carrier offset, its argument was
unwrapped and linearly interpolated to the native sample times; the carrier
$2\pi f_0t$ was then restored. The phase convention and frequency bands were
specified before the band-power and higher-order statistics were evaluated.

Within each geometry, samples from the other records were accumulated in
256 equal phase bins. Each bin mean was the pooled sample sum divided by the
pooled sample count. The Fourier series of these means retained harmonics
satisfying $nf_0\leq3$~MHz, where $f_0$ was the held-out record's
breathing-frequency estimate. The truncated series was evaluated on a
4096-point phase grid, and periodic linear interpolation supplied the
waveform along the held-out phase trajectory. The nominal 3~MHz limit was
chosen to cover all residual-power bands, including 1--3~MHz;
it is a template harmonic-order cutoff, not a temporal low-pass filter
applied after waveform reconstruction. The fixed-probe decomposition supplies
the component signals used in
the auto-bicoherence and signed third-order analyses below, whereas the
movable-probe signal supplies the second channel for the two-probe regression.
A reciprocal control repeats the complete decomposition using the
movable-probe phase $\phi_Y$ to condition both signals.

A step-by-step explanation and an illustrative Python implementation of
the waveform-extraction procedure are provided in the K04 module of
AlgoPlasma.\cite{algoplasma2026}

Figure~\ref{fig:waveform-residual-example} illustrates this decomposition
for a held-out fixed-probe record in both time and frequency. The other nine
records from the same probe-pair configuration supplied the template. The repeatable
waveform follows the dominant cyclic variation, whereas the residual retains
smaller irregular fluctuations. In the \mbox{0.5--1~MHz} band of this record, the
residual spectrum nearly coincides with the measured spectrum and lies well
above the waveform spectrum.

This is a repeatability-based decomposition rather than a low-/high-frequency
split. Because the waveform retains breathing harmonics through
3~MHz, $C$ can contain higher-frequency harmonic content, while $R$ can retain
power whose intensity varies with breathing phase. The residual therefore
denotes only what remains after subtraction of the cross-fitted conditional
mean; it is not identified a priori with turbulence or a distinct plasma mode.
This finite-record construction does not require the held-out residual's
band-limited conditional mean to vanish.
Its relation to theoretical cyclostationary decompositions is discussed in
the supplementary material, Sec.~S2.

\subsection{Power identity and phase-resolved residual power}

Two second-order quantities were evaluated from the cross-fitted
decomposition: the spectral contributions of $C$ and $R$, and the variation
of residual power with breathing phase. At each frequency, the decomposition
obeys the identity
\begin{equation}
 P_X(f)=P_C(f)+P_R(f)+2\operatorname{Re}S_{CR}(f).
 \label{eq:power-identity}
\end{equation}
Here, $P_U$ denotes the spectral power of $U$, and $S_{CR}$ is the
cross-spectrum between the fixed-probe waveform and residual. For each record
and band, the three right-hand-side terms were integrated and normalized by
the corresponding measured-signal power. The resulting ratios were averaged
with equal record weights within each geometry. Figures and tables report
separate medians of the 20 geometry summaries, so their displayed values need
not sum exactly to unity even though the identity closes for every record and
band. The phase-resolved numerical closure check and its error are reported
in the supplementary material, Sec.~S3.

To determine whether residual power remained organized by breathing phase,
we defined its phase-resolved value and peak-to-mean modulation in band $B$ as
\begin{equation}
 \begin{aligned}
 P_{R,B}(\phi)&=\mathbb{E}\!\left[|z_{R,B}(t)|^2
 \mid\phi(t)=\phi\right],\\
 M_B&=\frac{\max_\phi P_{R,B}(\phi)}
 {\langle P_{R,B}\rangle_\phi},
 \end{aligned}
 \label{eq:phase-power}
\end{equation}
where $z_{R,B}$ is the band-limited complex envelope. This quantity is a
conditional second moment. It is not assumed to equal conditional variance,
which would additionally subtract
$|\mathbb{E}[z_{R,B}\mid\phi]|^2$. For each record, $|z_{R,B}|^2$ was
averaged in 96 equal phase bins, and the resulting curve was normalized by its
phase mean. The normalized curves were averaged with equal record weights
within each geometry, after which $M_B$ was calculated from the aggregate
curve. Thus, $M_B$ is not an average or median of record-level peak-to-mean
ratios. No additional smoothing was applied to the phase-power curves.

The circular-shift test examined whether the record-level power profiles
remained aligned at a common breathing phase after geometry aggregation. For
each record, the residual-power samples were unchanged, while their
correspondence to the phase trajectory was shifted before rebinning and
normalization. Each of 2,000 null draws independently selected one of 64
shifted curves for every record and then repeated the observed geometry-level
averaging and peak extraction. The geometry-specific threshold was the 95th
percentile of the resulting $M_B$ distribution. The displacement range and
randomization sequence are specified in the supplementary material, Sec.~S3.

\subsection{Bicoherence comparison under a common estimator}

Unlike band power, bicoherence does not provide an additive component budget.
We therefore asked whether the harmonic phase organization of $X$ was
reproduced when only its waveform or residual was retained.
For $U\in\{X,C_X,R_X\}$, auto-bicoherence was evaluated with identical
segmentation and frequency grid:
\begin{equation}
 b_U^2(f_1,f_2)=
 \frac{\left|\left\langle U(f_1)U(f_2)U^*(f_1+f_2)\right\rangle\right|^2}
 {\left\langle|U(f_1)U(f_2)|^2\right\rangle
  \left\langle|U(f_1+f_2)|^2\right\rangle}.
 \label{eq:bicoherence}
\end{equation}
Angle brackets denote averaging over the non-overlapping 10,000-sample
Hann-windowed segments used throughout the higher-order analysis.
Each signal used its own normalization denominator in
Eq.~\eqref{eq:bicoherence}; averaging over the common region of interest
(ROI) used weights derived from the $X$ denominator.

The signed input grid comprised nonzero 10~kHz multiples satisfying
$0<|f_1|,|f_2|\leq1.5$~MHz, with 0.01~MHz as the first nonzero input.
Retained triads also required a positive sum-frequency leg satisfying
$0.05\leq f_1+f_2\leq1.5$~MHz. Negative-frequency coefficients were obtained
by complex conjugation, and each unordered input pair was evaluated once.
The resulting values were displayed at both $(f_1,f_2)$ and $(f_2,f_1)$.

For each record, the ROI was specified without using any
held-out bicoherence value. It comprised integer sum- and difference-frequency
triads whose positive target leg spanned the second through twelfth harmonics
of the measured $f_0$. All three frequency legs also had to exceed $-60$~dB
relative to the maximum $X$ power over 0.05--1.5~MHz. The principal
record-level summary was the $X$-denominator-weighted mean $b^2$ within this
ROI. The shared cells and $X$-derived weights provide an estimator-matched
comparison, not a decomposition into fractional bicoherence contributions.

Each residual was also compared with 200 Fourier-magnitude-preserving
random-phase surrogates \cite{theiler1992surrogate}. Each surrogate retained every segment-by-frequency magnitude while
independently randomizing the Fourier phases. This operation left the residual
bicoherence denominator of every ROI cell unchanged while removing coherent
Fourier-phase relations. The 95th percentile of the surrogate weighted-ROI
statistic defined a record-specific, prespecified random-phase baseline.
One-factor sensitivity checks varied segment count or placement, the common
power-support floor, and maximum harmonic order. The fast Fourier transform
(FFT) window length and overlap remained fixed; the tested settings are listed
in the supplementary material, Sec.~S4.

\subsection{Cross-fitted signed third-order functional}

Bicoherence shows whether the waveform or residual reproduces harmonic phase
organization, but its normalized squared magnitude does not admit an additive
component decomposition. We therefore introduce a complementary signed
third-order functional whose three signal inputs can be replaced by waveform
or residual components while the fitted operator is held fixed. At the level
of an individual triad, the expansion follows directly from
\begin{equation}
 \begin{aligned}
 \left\langle X_aX_bX_k^*\right\rangle
 &={}\left\langle(C_a+R_a)(C_b+R_b)(C_k+R_k)^*\right\rangle.
 \end{aligned}
 \label{eq:triad-input-expansion}
\end{equation}
The resulting eight products are grouped by the number of residual inputs
into $CCC$, one-$R$, two-$R$, and $RRR$ contributions.

We evaluate this expansion with a cross-fitted modified Ritz-type functional
in the nonlinear-transfer-function lineage of Ritz and Powers and subsequent
experimental implementations
\cite{ritz1986,ritz1989,kim1996,brown2023prl,brown2023pre}. It combines
training-only regressor screening, leave-one-record-out fitting, a fixed
target band, and an exact $C/R$ expansion. Only the three signal inputs change
during component attribution; the fitted weights, selected regressors,
cross-probe phase, target band, and normalization remain common to every
substitution. This common operator is what makes the grouped terms an exact
expansion of one specified statistic. Because no propagation-speed or
probe-separation factor is introduced, the normalized output is dimensionless
rather than a calibrated transfer rate or flux.

The two-probe regression estimates how quadratic combinations of the
fixed-probe Fourier components relate to the movable-probe response. The
fitted coefficients then weight the corresponding third-order moments
evaluated in the held-out record.

For held-out record $r$, the trilinear functional over band $B$ is
\begin{equation}
 \begin{split}
 \mathcal{G}_{B,r}(U,V,W)
 &=\sum_{k:f_k\in B}\frac{1}{2P_{X,r}(k)}
 \operatorname{Re}\!\Biggl[e^{-\mathrm{i}\Delta\theta_k^{(-r)}}
 \\
 &\qquad\times
 \sum_{(a,b)\in\mathcal{Q}_k^{(-r)}}Q_{ab,k}^{(-r)}
 \left\langle U_aV_bW_k^*\right\rangle_r\Biggr].
\end{split}
\label{eq:signed-functional}
\end{equation}
Here, $U$, $V$, and $W$ denote the three input signals. For component
attribution, each input is chosen as $C_X$ or $R_X$. The nonzero signed input
bins obey $a+b=k$ with $|f_a|,|f_b|\leq1.5$~MHz. The coefficients $Q$ relate selected
quadratic products of the fixed-probe signal to the movable-probe response,
and $\Delta\theta_k^{(-r)}=\arg\langle Y_kX_k^*\rangle_{-r}$ is the
training-set cross-probe phase at the target bin. For each fold, $Q$,
$\Delta\theta$, and the regressor set $\mathcal{Q}$ are estimated from the
other complete records in the same geometry.

Candidate quadratic products were retained when their training-set
auto-bicoherence exceeded
$\max[0.05,-\ln(0.01)/N_{\mathrm{train}}]$, where $N_{\mathrm{train}}$ is the
number of training segments, and at most the 32 highest-scoring candidates
were used. Before fitting, the linear projection of $Y_k$ on $X_k$ was removed
from the movable-probe target, and each quadratic regressor was projected off
$X_k$. The primary and strict-sensitivity fits used relative singular-value
cutoffs of $10^{-3}$ and $10^{-1}$, respectively. No held-out bicoherence
value entered the selection or fit. Non-overlapping 10,000-sample Hann
segments at 100~MS/s give 10~kHz spacing. The primary band comprises four bin
centers at 50, 60, 70, and 80~kHz.

Within each held-out fold, the coefficients $Q$, cross-probe phases
$\Delta\theta$, selected regressor set $\mathcal{Q}$, target band, and
normalizing spectrum $P_{X,r}$ are held fixed across all component
substitutions. The normalizing spectrum
$P_{X,r}(k)=\langle|X_k|^2\rangle_r$ is computed from the full measured signal
$X$ in record $r$, rather than separately from each substituted component.
The regressors are not rescreened, $Q$ is not refitted, and the denominator is
not changed for individual terms. For a fixed target band $B$, we denote the
total output $\mathcal{G}_{B,r}(X,X,X)$ by $S_{\mathrm{total},r}$. With the record index suppressed for
brevity, substituting $X=C_X+R_X$ into the three inputs gives the exact
expansion
\begin{equation}
 \mathcal{G}_B(X,X,X)=
 S_{CCC}+S_{1R}+S_{2R}+S_{RRR}.
 \label{eq:third-order-expansion}
\end{equation}
Here $S_{CCC}$ is the all-waveform contribution, with three $C_X$ inputs;
$S_{1R}$ and $S_{2R}$ are the one-residual and two-residual contributions,
each summing the three corresponding input permutations; and $S_{RRR}$ is the
all-residual contribution. The terms are signed and can cancel, so their
ratios are algebraic comparisons rather than mutually exclusive fractions of
conserved energy. Further operator and expansion details are provided in the
supplementary material, Sec.~S5.

\FloatBarrier

% ============================================================
\section{Results}
\label{sec:results}
% ============================================================

The results address two related but distinct questions: which component
carries the measured power above 0.2~MHz, and which component reproduces
higher-order phase organization at the breathing harmonics. We first
establish the breathing-frequency range, quantify band-power allocation, and
determine whether residual power remains organized by breathing phase. We then
compare bicoherence over the prescribed harmonic region and decompose a signed
third-order statistic over 50--80~kHz near the second breathing harmonic.
Phase-reference and acquisition-order checks assess the sensitivity of these
attributions.

\subsection{Record-level breathing frequency}

The record-level breathing frequencies provided the reference for locating
the harmonics considered below.
Across the \FreezeFZeroRecordCount{} complete records, the estimated breathing
frequencies from the fixed-probe signals clustered near 39.3~kHz
(supplementary material, Fig.~S2(b)). Their mean was
\FreezeFZeroRecordMeanKHz~kHz, their median was
\FreezeFZeroRecordMedianKHz~kHz, and their range was
\FreezeFZeroRecordMinKHz--\FreezeFZeroRecordMaxKHz~kHz. This range describes
variation across records. The waveform--residual decomposition used each
record's own breathing-phase trajectory.

\subsection{Residual-dominated power in the 0.2--1~MHz bands}

\begin{figure*}[!t]
  \centering
  \includegraphics[width=\figwideone]{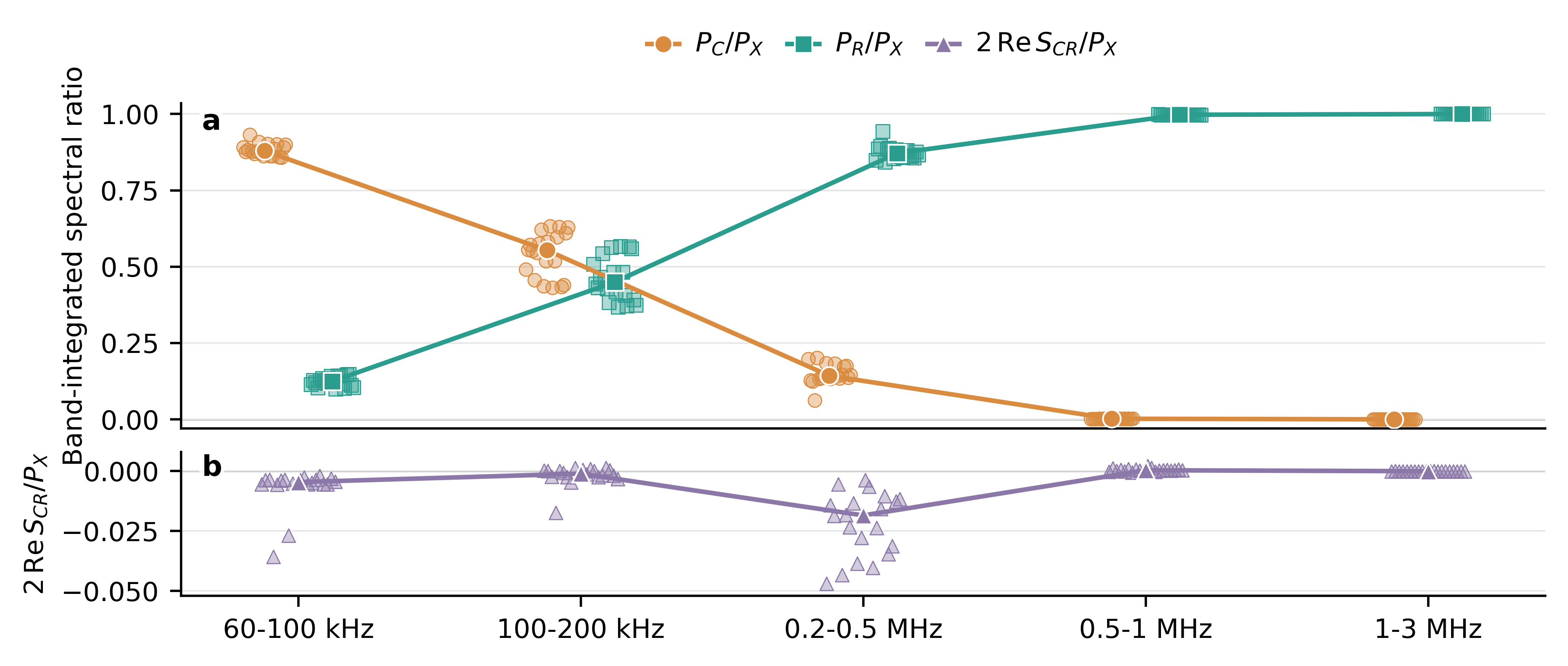}
  \caption{Five-band power attribution to the waveform and
  residual across 20 geometries. (a) Orange circles and green squares show
  $P_C/P_X$ and $P_R/P_X$ on a common ordinate. (b) Purple triangles show
  $2\operatorname{Re}S_{CR}/P_X$ on an expanded ordinate. Solid lines connect
  componentwise medians of the 20 geometry summaries; semi-transparent
  outlined symbols show individual geometry values with horizontal offsets
  only.
  Because the displayed component medians are taken
  separately, they need not sum exactly to unity.}
  \label{fig:frequency-crossover}
\end{figure*}

\begin{table}[!t]
\caption{Band-integrated phase-locked, residual, and cross-term power with
phase-resolved residual modulation. Entries are componentwise medians over 20
geometry summaries; ``above threshold'' is the number of geometries above the
95th percentile of the circular-shift distribution.}
\label{tab:power-attribution}
\centering
\resizebox{\columnwidth}{!}{%
\begin{tabular}{@{}lrrrrr@{}}
\hline\hline
Band & $P_C/P_X$ & $P_R/P_X$ & \shortstack{$2\mathrm{Re}S_{CR}$\\$/P_X$} & $M_B$ & \shortstack{Above shift\\threshold} \\
\hline
\FreezePowerTableRows
\hline\hline
\end{tabular}
}
\end{table}

We first assessed how much of the higher-frequency power in the plume signals
was accounted for by the repeatable breathing waveform. The decomposition
retained waveform harmonics through 3~MHz, so higher-frequency content was
not assigned to the residual by construction.

The waveform dominated the 60--100~kHz band, while the waveform and residual
contributions became comparable at 100--200~kHz
(Fig.~\ref{fig:frequency-crossover}(a)). At higher frequencies, the residual
carried approximately 87.1\% and 99.7\% of the measured power in the
0.2--0.5 and 0.5--1~MHz bands, respectively. These percentages are medians
across the 20 geometry summaries; Table~\ref{tab:power-attribution} reports
the componentwise medians for all five bands.

The cross term required by Eq.~\eqref{eq:power-identity} remained small in
the geometry-median summaries. Its largest normalized magnitude was 0.0185,
in the 0.2--0.5~MHz band (Fig.~\ref{fig:frequency-crossover}(b)). Negative
values represent partial cancellation between the waveform and residual
contributions to measured power, rather than negative component power.
The much larger opposing changes in $P_C/P_X$ and $P_R/P_X$ therefore
reflect a shift in which component carries the power, rather than a change
dominated by the cross term.

Thus, the repeatable breathing waveform did not account for most of the
measured power in the 0.2--1~MHz bands. Residual dominance does not, however,
establish independence from the breathing oscillation. We next examined
whether the power remaining after waveform removal still varied over the
breathing cycle.

\subsection{Breathing-phase modulation of residual power}

\begin{figure*}[!t]
  \centering
  \includegraphics[width=\figwideone]{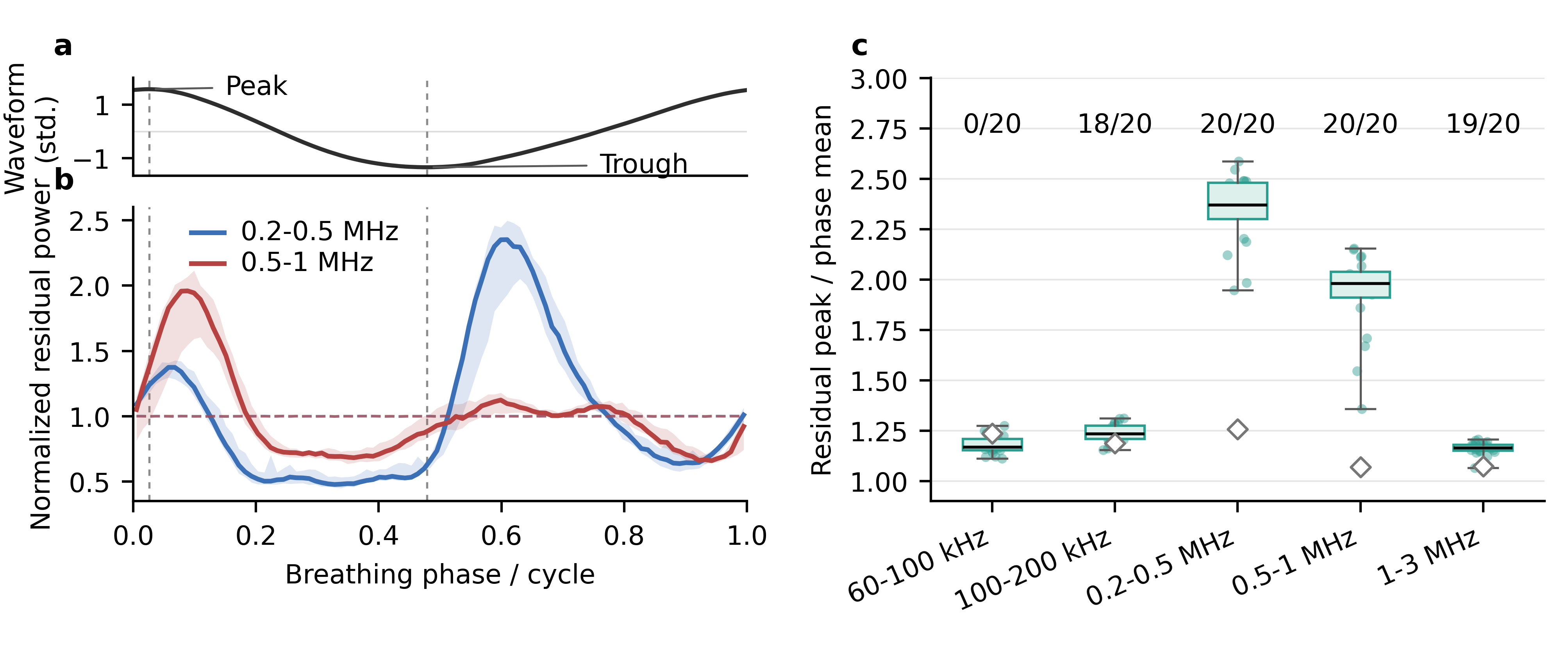}
  \caption{Phase-resolved residual power and circular-shift comparison.
  (a) Standardized phase-folded fixed-probe waveform, shown as the
  cross-geometry median with 10th--90th percentile shading; vertical dashed
  lines mark its maximum and minimum under the original phase convention.
  (b) Solid lines show cross-geometry medians of the residual-power profiles
  in 0.2--0.5 and 0.5--1~MHz. Each record's profile is divided by its own
  unweighted phase-bin mean before equal-record averaging within each
  geometry; shading spans the 10th--90th percentiles across geometry summaries.
  The normalized profiles compare phase dependence, not absolute band power.
  Colored dashed lines show cross-geometry
  medians after averaging all shifted curves within each geometry; these
  display curves do not define the thresholds in panel (c). (c) Points show
  geometry-level peak-to-mean ratios, with horizontal jitter for visibility.
  Boxes span the interquartile range, with median lines and full-range
  whiskers. Open diamonds show the cross-geometry median of the
  geometry-specific 95th-percentile circular-shift thresholds. Counts compare
  each geometry's observed statistic with its own threshold.}
  \label{fig:residual-phase-gating}
\end{figure*}

Residual power retained reproducible organization with breathing phase in the
two sub-MHz bands after removal of the repeatable waveform. To describe its
variation within the breathing cycle, we used the standardized fixed-probe
waveform as a phase reference. Its cross-geometry median peaked near
0.03 cycle and reached a minimum near 0.48 cycle
(Fig.~\ref{fig:residual-phase-gating}(a)).

The two bands had different dominant peak positions in their cross-geometry
median power profiles (Fig.~\ref{fig:residual-phase-gating}(b)). For
0.5--1~MHz, the dominant maximum lay near 0.1 cycle, on the descending side
of the probe-waveform peak. For 0.2--0.5~MHz, it lay near 0.6 cycle, during
recovery from the waveform minimum. These phases describe features of the
median profiles relative to the measured probe waveform; they do not assign
the power maxima to specific ionization or acceleration stages. Because
each record's profile was normalized by its own phase mean before geometry
averaging, the plotted heights describe relative modulation within each band,
not the absolute power difference between bands.

The geometry-median peak-to-mean ratios $M_B$ were 2.370 for 0.2--0.5~MHz
and 1.981 for 0.5--1~MHz. In both bands, $M_B$ exceeded the geometry-specific
95th-percentile circular-shift threshold in all 20 geometries
(Fig.~\ref{fig:residual-phase-gating}(c)); Table~\ref{tab:power-attribution}
gives the corresponding counts for the other bands. This comparison supports
reproducible phase--power alignment within each band under the existing
decomposition; it does not test the difference between the two peak phases.
Thus, removing the repeatable breathing waveform left most of the sub-MHz
power in the residual while retaining its organization within the breathing
cycle.

\subsection{The waveform reproduces the principal breathing-harmonic bicoherence}

\begin{figure*}[!t]
  \centering
  \includegraphics[width=\figwideone]{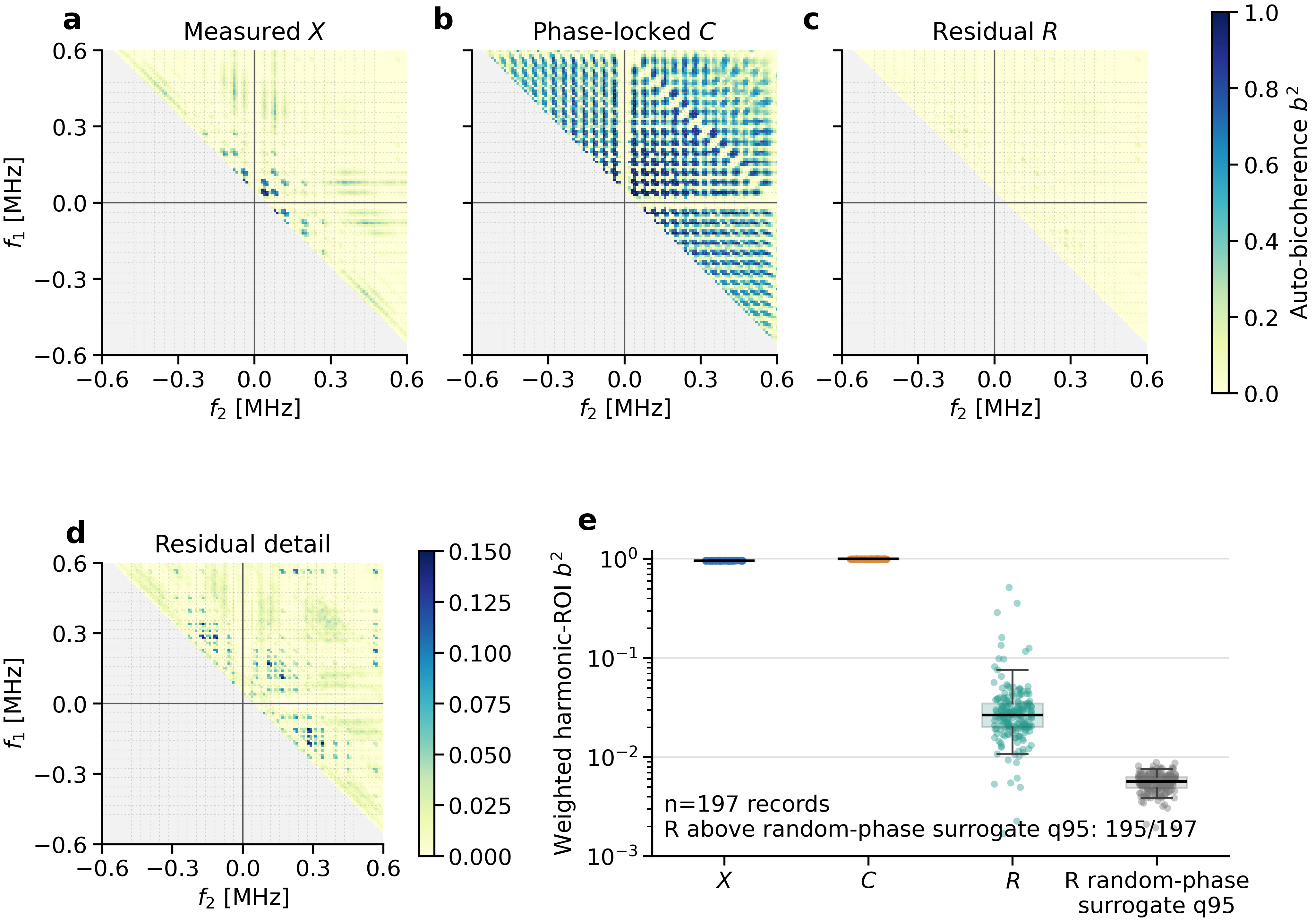}
  \caption{Component-resolved bicoherence maps and harmonic-ROI summaries.
  (a--c) Auto-bicoherence of the measured signal $X$, waveform $C_X$, and
  residual $R_X$ for the probe-pair configuration illustrated in
  Fig.~\ref{fig:waveform-residual-example},
  evaluated with the same estimator and displayed on the common
  $X$-power-supported frequency plane with a shared 0--1 color scale.
  These panels show the central $\pm0.6$~MHz region. (d) Residual detail on
  the same frequency support, using an
  independent 0--0.15 color scale, saturated above 0.15 and not directly
  comparable with panels (a--c). (e)
  Record-level distributions of weighted harmonic-ROI bicoherence for $X$, $C_X$, and
  $R_X$, together with the record-specific random-phase surrogate
  95th-percentile thresholds, across all 197 complete records. Points denote
  records, and boxes show interquartile ranges with median lines and
  5th--95th-percentile whiskers; the ordinate is logarithmic. The ROI summaries
  use the shared $X$-power support and $X$-derived denominator weights.}
  \label{fig:bicoherence-attribution}
\end{figure*}

Whereas the power identity assigns measured band power among additive terms,
the bicoherence comparison asks whether the strong breathing-harmonic
signature persists in the residual after waveform removal. On the common
0--1 scale for the illustrated configuration, the principal harmonic features
of $X$ also appeared in the
waveform $C_X$, whereas the residual $R_X$ was substantially weaker
(Fig.~\ref{fig:bicoherence-attribution}(a--c)). The expanded residual-only
scale revealed weaker localized structure within the same support
(Fig.~\ref{fig:bicoherence-attribution}(d)).

Across all
\FreezeResidualNullTotal{} complete records, the record-level medians of the
weighted harmonic-ROI statistic were \FreezeBicoXRecordMedian{} for $X$,
\FreezeBicoCRecordMedian{} for $C_X$, and \FreezeBicoRRecordMedian{} for
$R_X$ (Fig.~\ref{fig:bicoherence-attribution}(e)). The ordering
$C_X>X>R_X$ also held for all 20 geometry summaries. These separately
normalized bicoherences compare phase organization over a common ROI;
they are not additive fractions of the measured bicoherence. Thus, the
waveform reproduced the strong harmonic signature even though the residual
carried most of the power in the reported sub-MHz bands.

Despite its smaller magnitude, the residual statistic exceeded its
record-specific 95th-percentile random-phase surrogate threshold in
\FreezeResidualNullCount/\FreezeResidualNullTotal{} records under the
prespecified baseline. This exceedance supports residual Fourier-phase
organization relative to the specified surrogate comparison, but does not
identify a nonlinear interaction among distinct plasma modes. The tested
estimator and ROI variants retained the separation between waveform and
residual bicoherence; their quantitative ranges
and the complete residual maps are provided in the supplementary material,
Sec.~S4. We next used an exactly expanded signed third-order statistic to
examine component contributions near the second breathing harmonic.

\begin{figure*}[!t]
  \centering
  \includegraphics[width=\figwideone]{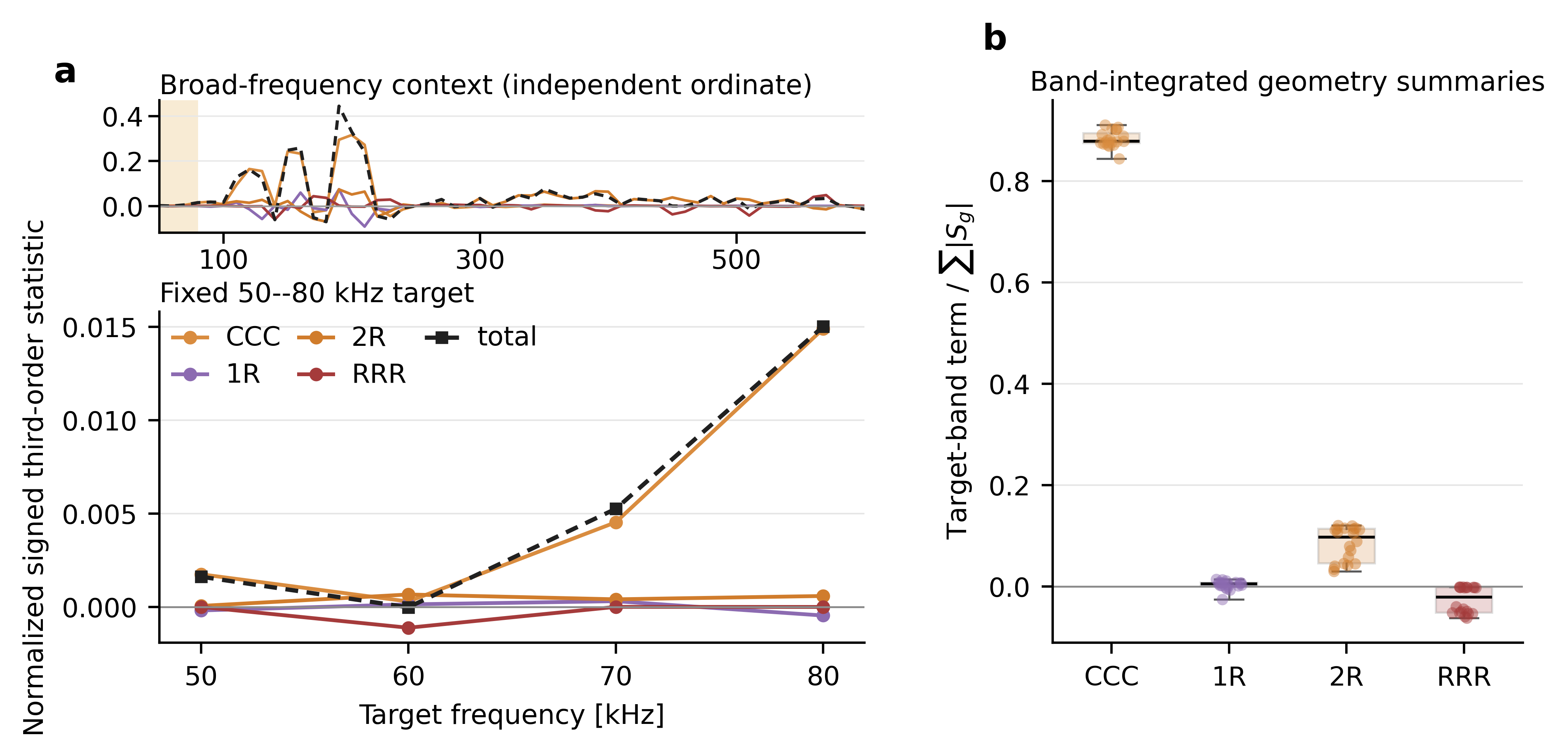}
  \caption{Exact component expansion of the target signed third-order
  statistic near the second breathing harmonic. (a) Frequency-resolved total,
  all-waveform ($CCC$), one-residual (one-$R$), two-residual (two-$R$), and
  all-residual ($RRR$) contributions for the probe-pair configuration
  illustrated in Fig.~\ref{fig:waveform-residual-example}. The lower axis enlarges
  the four bins in the fixed 50--80~kHz target; the upper axis retains the
  broader frequency range on an independent ordinate, with the target shaded.
  The broader trace provides spectral context only. (b) Primary-cutoff, band-integrated
  component summaries across 20 geometries. Within each geometry, each
  component is divided by the sum of the absolute values of all four component
  groups, rather than the algebraic total used in
  Fig.~\ref{fig:third-order-robustness}(a). Points denote geometries; boxes show interquartile ranges with median
  lines and full-range whiskers.}
  \label{fig:third-order-components}
\end{figure*}

\begin{figure*}[!t]
  \centering
  \includegraphics[width=\figwideone]{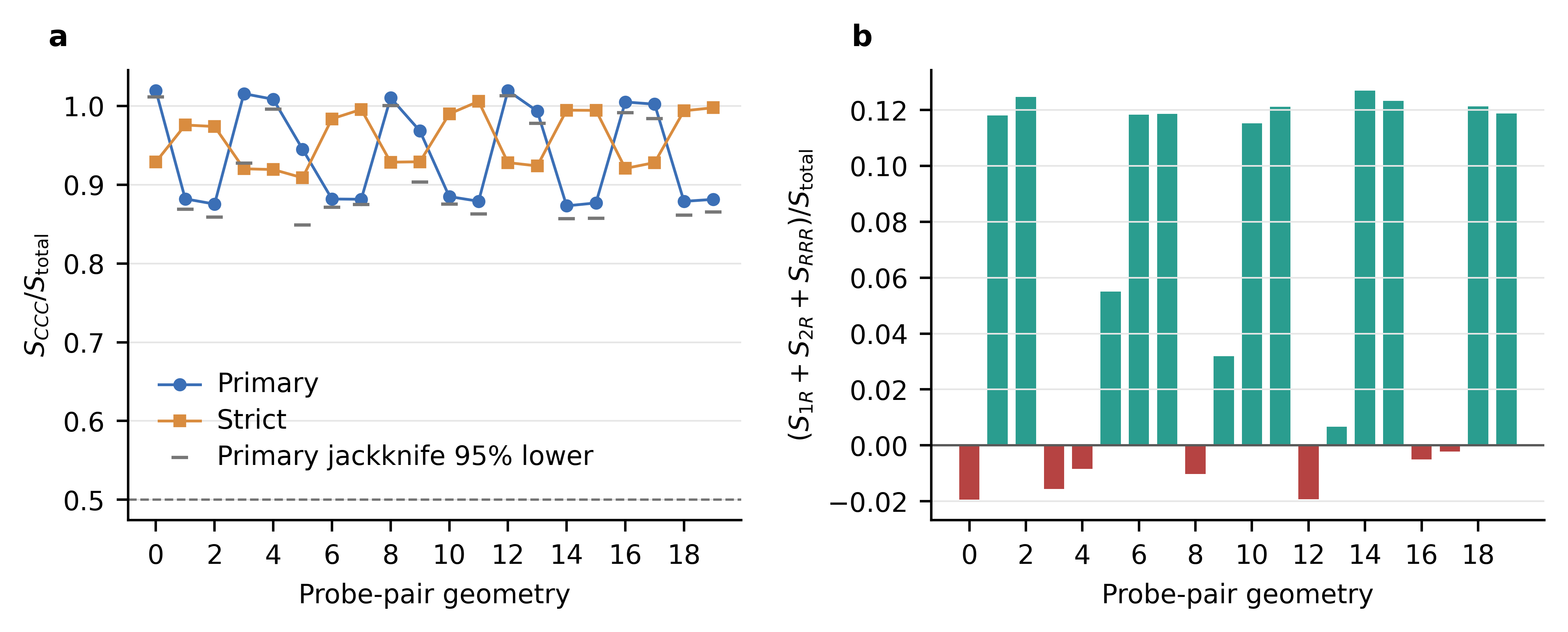}
  \caption{Robustness and boundary of the signed third-order attribution.
  (a) $S_{CCC}/S_{\mathrm{total}}$ for the primary and strict relative
  singular-value cutoffs across 20 geometries. Horizontal marks show the lower
  endpoints of the primary two-sided 95\% full-refit jackknife intervals
  for this ratio; the
  breathing-phase trajectories were held fixed during refitting, and the
  dashed line is an algebraic 0.5 reference. (b) Primary-cutoff, equal-record
  normalized residual-associated sum
  $(S_{1R}+S_{2R}+S_{RRR})/S_{\mathrm{total}}$ for each geometry.
  Both panels use the algebraic total, which includes cancellation among
  signed terms; the ratios are not energy fractions.}
  \label{fig:third-order-robustness}
\end{figure*}

\subsection{Exact third-order attribution near the second breathing harmonic}
\label{sec:third-order-results}

We used the fixed 50--80~kHz target near the second breathing harmonic to
test whether waveform dominance persisted when mixed and all-residual terms
were included explicitly in an additive signed statistic.
The band-integrated statistic was dominated by the all-waveform contribution
$S_{CCC}$ across the 20 geometries
(Fig.~\ref{fig:third-order-components}). For the primary $10^{-3}$ and strict
$10^{-1}$ relative
singular-value cutoffs, the geometry-median $CCC/\mathrm{total}$ ratios were
approximately 0.915 and 0.951, respectively. All 20 geometries exceeded the
algebraic 0.5 reference under both cutoffs
(Fig.~\ref{fig:third-order-robustness}(a)).

This attribution applies to a fixed harmonic neighborhood. The target
comprised \FreezeTargetBinCount{} bin centers at 50, 60, 70, and 80~kHz,
with 10~kHz spacing. Across records, $2f_0$ ranged from
\FreezeTwoFZeroRecordMinKHz{} to \FreezeTwoFZeroRecordMaxKHz{}~kHz and crossed
the upper band edge for some records; the target did not track each record's
$2f_0$.

The signed components can reinforce or cancel one another, as illustrated
by the individual target bins in Fig.~\ref{fig:third-order-components}(a).
The component comparison uses their band-integrated values:
Fig.~\ref{fig:third-order-components}(b) divides each component by the sum
of the absolute component magnitudes, whereas
Fig.~\ref{fig:third-order-robustness}(a) divides $S_{CCC}$ by the algebraic
total. Cancellation in that total allows $CCC/\mathrm{total}$ to exceed
unity. Neither normalization represents fractions of physical energy.

With the breathing-phase trajectories held fixed, the full-refit jackknife
\cite{miller1974jackknife} rebuilt the waveform--residual decomposition and
refitted the third-order operator. For the primary cutoff, the lower endpoints
of the two-sided 95\% intervals for $CCC/\mathrm{total}$ exceeded 0.5 in all
20 geometries (Fig.~\ref{fig:third-order-robustness}(a)). Excluding the bin nearest
record-wise $2f_0$ and shifting the target to 40--80 or 50--90~kHz also
retained the attribution (supplementary material, Sec.~S5.2).
These checks support the attribution within the specified
target and tested neighboring bands, rather than across all harmonics.

The smaller residual-associated contributions had mixed signs. For the
primary cutoff, the equal-record normalized sum was positive in
\FreezeSignedPrimaryResidualPositiveCount/20 geometries
(Fig.~\ref{fig:third-order-robustness}(b)). By contrast, conditional
record-bootstrap intervals \cite{efron1979bootstrap} for the record-amplitude-weighted unnormalized
aggregate were entirely negative in 18/20 geometries and crossed zero in the
other two. The sign of the aggregate residual-associated contribution
therefore depended on the weighting and normalization convention. Without
a calibrated propagation factor, neither sign identifies an energy-transfer
direction.

An envelope-preserving Gaussian null approximately matched phase-binned
residual second moments and residual auto- and cross-spectra. The primary-cutoff,
equal-record mean of the normalized residual-associated signed statistic,
without division by $S_{\mathrm{total}}$, exceeded its one-sided
95th-percentile threshold in \FreezeEnvelopeNullCount/\FreezeEnvelopeNullTotal{}
geometries (supplementary material, Sec.~S5). This comparison supports residual
third-order structure beyond that reproduced by the specified construction, without
identifying its physical origin.

\subsection{Sensitivity to phase reference and acquisition order}

We tested whether power allocation and residual-power modulation depended on
which probe supplied the breathing phase. For each target signal, the
same-reference analysis used its own phase and the cross-reference analysis
used the other probe's phase. In the two sub-MHz bands, the absolute paired
median changes in waveform and residual power fractions were below
$6\times10^{-4}$, although their geometry-cluster bootstrap 95\% intervals
excluded zero. For residual peak-to-mean ratios, all four paired-difference
intervals across the two targets and two bands included zero. After aligning
the phase origins between references, median absolute peak-phase displacements
were 0.0055--0.0057 breathing cycles (supplementary material, Sec.~S7).

For the fixed-probe signal, the record-level median residual bicoherence
increased from 0.02654 to 0.02866 under the cross-reference analysis,
with a paired median difference of 0.00280 and a geometry-cluster bootstrap
95\% interval of $[0.00079,0.00402]$. The paired estimate is the median of
record-wise cross-reference minus same-reference differences.
The corresponding movable-probe change
was smaller and its interval included zero (supplementary material, Sec.~S7).
Residual bicoherence nevertheless remained weak relative to the
measured-signal and waveform values under both phase references. The principal
scale separation was therefore retained, while the fixed-probe shift defined
a local phase-reference sensitivity.

Without absolute timestamps, we used the file identifier as a within-geometry
acquisition-order proxy. After normalizing order within each geometry and
removing geometry-specific intercepts, the slopes for $f_0$, the five residual
power fractions, residual bicoherence, and $CCC/\mathrm{total}$ all had
geometry-cluster bootstrap 95\% intervals containing zero (supplementary
material, Sec.~S7). No common linear within-geometry order trend was detected
for these outcomes. This check does not exclude time-dependent drift or
establish simultaneous spatial structure across the sequentially acquired
geometries.

% ============================================================
\section{Discussion}
\label{sec:discussion}
% ============================================================

The discussion considers why power allocation and higher-order phase
organization must be distinguished, how the component comparisons constrain
spectral interpretation, and where the measurement and statistical boundaries lie.

\subsection{Power allocation and higher-order phase organization}

Power allocation and harmonic phase organization describe different aspects
of the recorded fluctuations. Residual dominance in the 0.2--0.5 and
0.5--1~MHz power bands does not imply a residual origin for the principal
breathing-harmonic bicoherence, which the repeatable waveform reproduces.
The waveform also dominates the specified signed third-order statistic in
the fixed 50--80~kHz target near the second harmonic. These comparisons concern
different statistics and specified frequency ranges; they are not
interchangeable measures of fluctuation strength.

Subtracting the estimated phase-conditioned mean does not require residual
band power to be independent of breathing phase. Cyclostationary analysis
likewise permits a periodic mean and periodically varying fluctuation
statistics to be separate features \cite{antoni2009}. Here, the measured
phase-dependent band power is a conditional second moment, not assumed to
equal conditional variance. Its different dominant maxima in the
cross-geometry median profiles suggest frequency-dependent organization
within the breathing cycle. The circular-shift comparison supports phase
modulation within each band, but does not establish a statistically
significant interband peak-phase difference. The peak positions neither
identify distinct physical modes nor locate specific ionization or
acceleration stages.

The waveform-dominated bicoherence is consistent with earlier observations of
fundamental--harmonic bicoherence accompanying a non-sinusoidal plasma waveform
\cite{kobayashi2012}. The waveform can itself reflect the dynamics of the
breathing cycle. Comparison under a common estimator shows that it reproduces
the principal harmonic signature, without identifying its physical origin
or classifying it as an artifact.

Frequency selection alone does not establish a residual origin when
harmonics of the repeatable waveform occupy the selected band. The
repeatable-waveform contribution should therefore be assessed before
harmonic bicoherence is attributed to interactions among residual
fluctuations.

\subsection{What the higher-order measures distinguish}

Bicoherence measures normalized third-order Fourier structure with amplitude
weighting, rather than phase relations alone. The waveform attribution rests
on comparison of the maps and weighted harmonic-ROI summaries under a common
estimator, not on a high waveform bicoherence value alone. The residual
statistic exceeds the specified random-phase baseline in most records, but
remains much smaller than those of $X$ and $C_X$. This residual structure
does not by itself isolate a nonlinear interaction \cite{aru2015}.

The random-phase surrogate comparison and envelope-preserving Gaussian null
address different questions. Random-phase surrogates preserve each
segment-by-frequency magnitude while removing coherent Fourier-phase
relations, providing a residual-bicoherence baseline conditional on the
observed amplitudes. The envelope-preserving null instead evaluates the
normalized residual-associated signed statistic in
Sec.~\ref{sec:third-order-results}. Two alternating rounds of phase-bin
rescaling and spectral rematching approximately reproduce residual second
moments while matching auto- and cross-spectra; they do not preserve the
pointwise envelope. Its constraints use each record's measured residuals,
so this is not a training-only prediction test (supplementary material,
Sec.~S5). Different constraints and outputs make these comparisons
complementary, not nested tests of increasing strength.

The signed functional adds an exact expansion of three signal inputs while
holding the fitted operator and normalization fixed. Its components can
cancel, so their ratios are algebraic contributions, not energy fractions.
The all-waveform contribution remained dominant across geometries, both
singular-value cutoffs, and the tested neighboring target bands. At the primary
cutoff, all geometry-level full-refit jackknife intervals for
$CCC/\mathrm{total}$ lay above 0.5 with the phase trajectories held fixed.
This supports attribution within the specified second-harmonic region, but
is not independent confirmation of the bicoherence result: regressor
screening uses training-side bicoherence. The aggregate residual-associated
sign also depends on the weighting and normalization convention. Together
with the absence of a calibrated propagation factor, this prevents its
interpretation as an energy-transfer direction.

The supplementary M0/M1 comparison has a separate scope: it retains the
waveform and generates Gaussian residuals with matched spectral covariance,
without phase gates (M0) or with bandwise amplitude gates (M1; supplementary
material, Sec.~S6). Neither construction jointly reproduced the evaluated
higher-order outputs. Incomplete reproduction of phase-dependent power,
differing calibration and evaluation bands, and indirect dependence through
precomputed templates limit this finding to the specified constructions.
It does not isolate residual non-Gaussianity or a plasma interaction
mechanism, and does not show that every second-order model is insufficient.

These comparisons constrain the statistical interpretation of the residual
features. A dynamical interpretation further requires assessing how
measurement response, waveform variability, and decomposition sensitivity
contribute to those features, as discussed below.

\subsection{Measurement scope and physical generality}

The conclusions apply to recorded probe-chain signals, not calibrated plasma
quantities or a physical transfer rate. The decoding and digital-range audits
establish the stored-voltage conversion and find no stored-word clipping;
they do not calibrate the analog response described in
Sec.~\ref{sec:experiment}. Instrumental nonlinearities can generate
higher-order spectral signatures \cite{walker2002}, but the present audits
neither establish nor exclude them. Residual structure may also reflect
physical waveform variability, phase error, and template mismatch. A
same-source two-channel bench test would help quantify the response of the
tested circuitry downstream of the injection point. It would not by itself
calibrate the probe--plasma response or identify the physical origin of
residual structure.

The physical generality is bounded by one operating condition and one axial
measurement station. The two channels were sampled simultaneously within
each record, but the 20 probe-pair geometries were acquired sequentially.
Agreement across geometries therefore supports reproducibility across
acquisitions, not simultaneous spatial structure over the full scan. The
absence of a detected common linear within-geometry acquisition-order trend
does not exclude time-dependent drift.

Calibration would constrain measurement-chain contributions, whereas
measurements at additional operating conditions and axial locations would
test the generality of the observed component attribution and residual-power
modulation. Identifying a microscopic mechanism would additionally require
observables that distinguish among candidate plasma processes.

% ============================================================
\section{Conclusions}
\label{sec:conclusions}
% ============================================================

In the paired Hall-thruster plume signals, the residual carried most of the
0.2--1~MHz power, whereas the repeatable waveform reproduced the principal
breathing-harmonic bicoherence. Residual band power nevertheless remained
reproducibly modulated by breathing phase after waveform removal.

For the specified signed third-order statistic over the fixed 50--80~kHz
target near the second breathing harmonic, the all-waveform term supplied
the dominant algebraic contribution. The geometry-median all-waveform-to-total
ratios were approximately 0.915 and 0.951 for the primary and strict
singular-value cutoffs, respectively; these are algebraic ratios, not energy
fractions. The residual retained weaker higher-order structure, while the
aggregate residual-associated sign depended on the weighting and
normalization convention. These results distinguish band-power allocation
from harmonic higher-order organization within the evaluated frequency ranges.

The diagnostic implication is to assess the repeatable-waveform contribution
to harmonic bicoherence and evaluate residual higher-order features with
statistic-matched controls before assigning a dynamical interpretation.
The conclusions apply to recorded probe-chain signals at one operating
condition and one axial measurement station, without analog-response
calibration. The component attribution does not identify a microscopic
interaction mechanism or measure a physical energy-transfer rate.

\section*{Supplementary Material}

See the supplementary material for the full decoding and digital-range audits;
phase construction and cross-fitting validation; power-identity closure and
circular-shift test; complete bicoherence maps, surrogate validation, and fixed-ROI
sensitivity; signed third-order functional details and robustness checks; M0/M1
construction and generated-output diagnostics; reciprocal phase-reference and
acquisition-order controls; and data provenance.

\section*{Acknowledgments}

This work was supported by the National Natural Science Foundation of China
(Grant Nos.~52472403 and U25B20235) and the State Key Laboratory of Advanced
Space Propulsion (Grant No.~2025ASP0306).

\section*{Author Declarations}

\subsection*{Conflict of Interest}

The authors have no conflicts to disclose.

\subsection*{Author Contributions}

\textbf{Zilong Peng}: Conceptualization (equal); Data curation (lead); Formal analysis (lead); Investigation (lead); Methodology (equal); Software (lead); Visualization (lead); Writing: original draft (lead).
\textbf{Baisheng Wang}: Resources (supporting); Investigation (supporting); Data curation (supporting); Validation (supporting).
\textbf{Zhongping Zhao}: Methodology (supporting); Software (supporting); Formal analysis (supporting).
\textbf{Yao Tang}: Investigation (supporting); Resources (supporting); Validation (supporting).
\textbf{Yinjian Zhao}: Conceptualization (equal); Funding acquisition (lead); Methodology (equal); Project administration (lead); Supervision (lead); Writing: review and editing (lead).
\textbf{Daren Yu}: Resources (supporting); Supervision (supporting); Writing: review and editing (supporting).
\textbf{Ming Zeng}: Resources (supporting); Supervision (supporting); Writing: review and editing (supporting).

\section*{Data Availability Statement}

The data that support the findings of this study are available from the corresponding author upon reasonable request.

% Tighten the gap before references to avoid a one-line continuation page.
\vspace{-8pt}
% Embedded references: compile without BibTeX or external bibliography files.
\makeatletter
% Skip REVTeX automatic .bbl insertion when using inline references.
\let\auto@bib@innerbib\@empty
\makeatother
%aipnum4-2.bst 2019-01-14 (MD) hand-edited version of apsrev4-1.bst
%Control: key (0)
%Control: author (72) initials jnrlst
%Control: editor formatted (1) identically to author
%Control: production of article title (1) required
%Control: page (1) range
%Control: year (1) truncated
%Control: production of eprint (0) enabled
%

% Balance the reference columns before starting the single-column supplement.
\onecolumngrid
\clearpage
\setcounter{section}{0}
\setcounter{subsection}{0}
\setcounter{figure}{0}
\setcounter{table}{0}
\setcounter{equation}{0}
\renewcommand{\thesection}{S\arabic{section}}
\renewcommand{\thesubsection}{\thesection.\arabic{subsection}}
\makeatletter
\renewcommand{\p@subsection}{}
\makeatother
\renewcommand{\thefigure}{S\arabic{figure}}
\renewcommand{\thetable}{S\arabic{table}}
\renewcommand{\theequation}{S\arabic{equation}}
\renewcommand{\theHsection}{si.\arabic{section}}
\renewcommand{\theHsubsection}{si.\arabic{section}.\arabic{subsection}}
\renewcommand{\theHfigure}{si.\arabic{figure}}
\renewcommand{\theHtable}{si.\arabic{table}}
\renewcommand{\theHequation}{si.\arabic{equation}}
\setlength{\emergencystretch}{2em}
% Author and affiliation details are shared with the manuscript title page.
\begin{center}
\LARGE\bfseries Supplementary Material
\end{center}
\vspace{1em}

\noindent\textbf{Scope and reading order.}
This Supplementary Material gives implementation details and validation for
the workflow in main-text Sec.~III~A and Fig.~3.
Sections~S1--S2 cover acquisition, decoding, phase extraction, and cross-fitted
decomposition; Sections~S3--S5 cover power, bicoherence, and signed third-order
attribution. Section~S6 contains the supplementary conditional Gaussian
comparison, including construction, constraint reproduction, higher-order
outputs, and interpretation limits. Section~S7 covers reciprocal phase
references and acquisition-order sensitivity.

The fixed- and movable-probe signals are denoted by $X(t)$ and $Y(t)$;
a frequency argument denotes Fourier coefficients. Both signals use the
fixed-probe phase $\phi_X$ for their channel-specific waveform--residual
decompositions; the reciprocal control uses $\phi_Y$.
Auto-bicoherence and the three inputs to the signed third-order functional use fixed-probe
components. The movable probe enters the training regression for the
quadratic-response coefficients $Q$ and the cross-probe phase (Section~S5).
The term \emph{deterministic} denotes
conditional-mean repeatability, not deterministic plasma dynamics.

% ------------------------------------------------------------
\section{Data acquisition, geometric calibration, and decoding}
\label{si:sec:decoding}
% ------------------------------------------------------------

\subsection{Data hierarchy and statistical units}

Table~\ref{si:tab:data-hierarchy} gives acquisition and conversion details;
main-text Sec.~II describes the operating point and probe arrangement.
Complete paired records are the analysis units. Geometry summaries and cluster
resampling assess reproducibility across sequential settings, not independent
experimental repetitions. Cycles, spectral segments, and frequency pixels
remain within-record quantities.

\begin{table}[H]
\centering
\caption{Acquisition and stored-word conversion for the paired probe records.}
\label{si:tab:data-hierarchy}
  \alttext{Acquisition summary for 197 paired records at 20 sequential probe geometries. Rows identify the simultaneous channels, bias and sensing circuit, binary encoding, sampling parameters, and voltage-conversion convention used throughout the analysis.}
\begin{tabularx}{\textwidth}{@{}>{\raggedright\arraybackslash}p{3.7cm}>{\raggedright\arraybackslash}X@{}}
\toprule
Item & Documented definition \\
\midrule
Dataset & 20 sequential geometries; 197 complete paired records
(8--10 per geometry), comprising 394 channel waveforms \\
Oscilloscope and channels & SDS814X-HD, 12-bit vertical resolution;
fixed probe on C1 and movable probe on C3, acquired simultaneously \\
Bias and sensing & HSPY-300-01 DC supply set to $-37$~V; voltage drops
across nominal 1~k$\Omega$ sensing resistors marked 50W1KJ \\
File format & Siglent V4; 4096-byte header followed by unsigned
16-bit little-endian waveform words \\
Sampling & 100~MS/s; $10^7$ samples (100~ms) per waveform \\
Voltage conversion & Per-file header conversion
$(u-32768)V_{\mathrm{div}}/N_{\mathrm{div}}+V_{\mathrm{offset}}$,
giving digitizer-input voltage \\
\bottomrule
\end{tabularx}
\end{table}

The probes were biased at $-37$~V, nominally in the ion-saturation regime.
Recorded voltages are relative ion-current fluctuation proxies, not calibrated
current or absolute plasma density. No contemporaneous current--voltage trace
or electron-temperature measurement was available.
The nominal 1~k$\Omega$ sensing resistance was identified from the 50W1KJ
component marking. Termination, intermediate analog components, usable
bandwidth, trigger setting, calibrated resistance, channel-dependent phase
response, and effective collection-area calibration were undocumented.
No post-acquisition amplitude- or phase-response correction was applied.
The statistics therefore describe the recorded probe-chain signals;
digital-range checks below do not exclude analog nonlinearities.

\subsection{Geometric calibration and sequential acquisition}

The four translation blocks were acquired sequentially, with rotation order
reversed between successive blocks rather than randomized.
Zero-translation tip spacings of 3.0, 2.9, and 2.7~mm at nominal
$0^{\circ}$, $45^{\circ}$, and $90^{\circ}$ determined the movable-stage
rotation center. We denote this center relative to the fixed probe by
$\bm C=(C_x,C_y)$. Selecting the solution branch satisfying $C_x<d_0/2$,
with $d_0$ the $0^{\circ}$ spacing, gave
$\bm C=(0.275,-0.036)$~mm.
A measured factor of 1.07 converted commanded translation to arm extension,
yielding separations of 2.7--4.6~mm (main-text Fig.~1(b)).
The three spacings and scale factor were held fixed; independent metrology
and their uncertainties were unavailable.

\subsection{Independent unsigned-integer decoding audit}

\begin{figure}[!htbp]
  \centering
  \includegraphics[width=\textwidth]{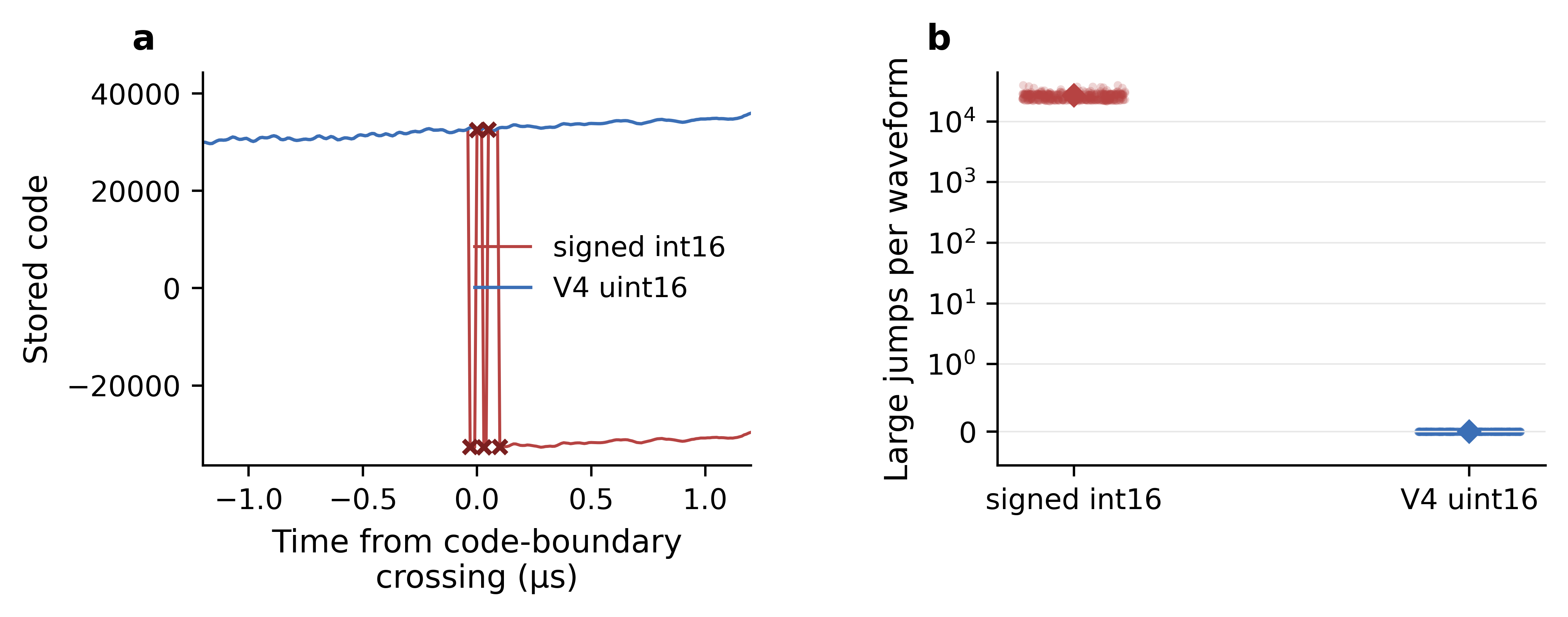}
  \caption{Binary-encoding audit. (a) The same representative waveform segment
  decoded as signed int16 and unsigned u2, with wrap events marked only for
  display. (b) Number of detected large jumps for all 394 record--channel
  waveforms. Points denote individual waveforms; diamonds and bars show
  medians and interquartile ranges.}
  \label{si:fig:encoding-audit}
  \alttext{Two panels compare decoding of stored probe signals. Signed decoding introduces large jumps at code-boundary crossings, whereas unsigned decoding removes these discontinuities. A distribution plot extends the comparison to all 394 channel waveforms.}
\end{figure}

The oscilloscope headers specified Siglent V4 records with 16-bit little-endian
waveform words, for which the file format prescribes unsigned-u2 decoding.
An independent audit compared signed-int16 and unsigned-u2 interpretations for
all 394 record--channel waveforms across 20 geometries
(Fig.~\ref{si:fig:encoding-audit}). Signed decoding produced 10,350,070 artificial
two's-complement wrap jumps; unsigned decoding produced none. The median ratio
of signed-decoded to u2-decoded power spectral density (PSD) was 59.877 in
1--3~MHz and 762.011 in 10--30~MHz. Record-level results are retained in the
detailed audit table underlying Fig.~\ref{si:fig:encoding-audit}. All analyses in
the main text and this Supplementary Material use unsigned-u2 decoding.

A separate range audit scanned the stored words in all 394 channel files
(Table~\ref{si:tab:instrument-range-audit}). No sample reached either digital
rail or lay within 64 codes of a rail; the longest exact-rail run was zero.
The probe-resolved headrooms likewise remained far from both rails. These
checks found no digital saturation in the stored words but do not test
analog-chain or other instrumental nonlinearities.

\FloatBarrier

\begin{table}[H]
  \centering
  \caption{Recorded-word range audit. Headrooms are the minimum values among
  the 197 files for each probe signal. ``Near rail'' denotes samples within
  64 codes of either digital rail. The audit concerns stored words only.}
  \label{si:tab:instrument-range-audit}
  \alttext{Digital-range checks for the fixed and movable probes list sample headroom, exact-rail files, and near-rail sample fractions. Neither channel reaches a digital rail; the checks do not assess the analog response.}
  \begin{tabularx}{\textwidth}{@{}>{\raggedright\arraybackslash}Xrrrrr@{}}
    \toprule
    Probe signal & Files & \shortstack{Minimum lower\\headroom (codes)} &
    \shortstack{Minimum upper\\headroom (codes)} &
    \shortstack{Files at an\\exact rail} &
    \shortstack{Maximum near-rail\\sample fraction} \\
    \midrule
    \SIInstrumentRangeRows
    \bottomrule
  \end{tabularx}
\end{table}
\FloatBarrier

% ------------------------------------------------------------
\section{Breathing phase and cross-fitted waveform--residual decomposition}
\label{si:sec:crossfit}
% ------------------------------------------------------------

\subsection{Breathing-phase coordinate}

For record $r$ in geometry $g$, the fixed-probe signal defined the primary
breathing phase $\phi_{X,g,r}(t)$. The complete demeaned record was Fourier
transformed, and the maximum-power discrete Fourier bin between 35 and 45~kHz
defined $f_0$. A rectangular 25--55~kHz positive-frequency band was transformed
to a complex baseband signal. After removal of the baseband carrier offset,
its argument was unwrapped and interpolated to the native 100~MS/s sample
times. The carrier $2\pi f_0t$ was then restored. This Fourier-band construction
required neither a finite-order causal filter nor an edge-discard rule. Phase
zero followed the complex Fourier convention and was not adjusted using a
target statistic. All phase-extraction settings were fixed before the
band-power and higher-order statistics were evaluated.
Figure~\ref{si:fig:phase-crossfit}(a) illustrates the measured signal and its
conditioning phase.

Across the \FreezeFZeroRecordCount{} complete records, the record-level $f_0$
estimates had a mean of \FreezeFZeroRecordMeanKHz~kHz, a median of
\FreezeFZeroRecordMedianKHz~kHz, and a range of
\FreezeFZeroRecordMinKHz--\FreezeFZeroRecordMaxKHz~kHz
(Fig.~\ref{si:fig:phase-crossfit}(b)). Main-text
Fig.~6(a) shows the cross-geometry median and 10th--90th percentiles of the
normalized phase-folded waveforms under this phase convention.

Training samples were accumulated in 256 equal phase bins, with each mean
formed from the pooled sample sum and count. The Fourier series of these
means retained harmonics satisfying $nf_0\leq3$~MHz, using the held-out
record's $f_0$. The truncated series was evaluated on a 4096-point phase
grid and periodically linearly interpolated along the held-out phase
trajectory. The nominal 3~MHz limit was chosen to cover every residual-power
band, including the 1--3~MHz calibration band; it limits template harmonic
order rather than applying a temporal low-pass filter to the reconstructed
waveform. Records required both probe signals, and each
geometry required at least three complete pairs. No additional phase-quality
exclusion was applied. Section~S7 tests sensitivity to replacing the primary
fixed-probe phase with the reciprocal movable-probe reference.

\subsection{Leave-one-record-out conditional mean}

For either channel $Z\in\{X,Y\}$, the held-out decomposition was
\begin{equation}
  Z_{g,r}(t)=C_{Z,g,-r}[\phi_{X,g,r}(t)]+R_{Z,g,r}(t).
  \label{si:eq:crossfit-decomposition}
\end{equation}
The waveform used all complete records at geometry $g$ except $r$.
The held-out record supplied its phase coordinate and the $f_0$ used for
harmonic selection, but no samples to the phase-binned training means;
subtraction defined $R_{Z,g,r}$.
The same components entered the power, bicoherence, and signed-third-order
analyses. The principal third-order operator also excluded $r$ during fitting
(Section~S5).

The residual can retain cycle-to-cycle variability, phase error, frequency
drift, template mismatch, instrument response, and independent higher-frequency
fluctuations. This finite-record, single-phase construction is not an
infinite-time cyclostationary projection, so the held-out residual's
band-limited conditional mean need not vanish.

\begin{figure}[!htbp]
  \centering
  \includegraphics[width=\textwidth]{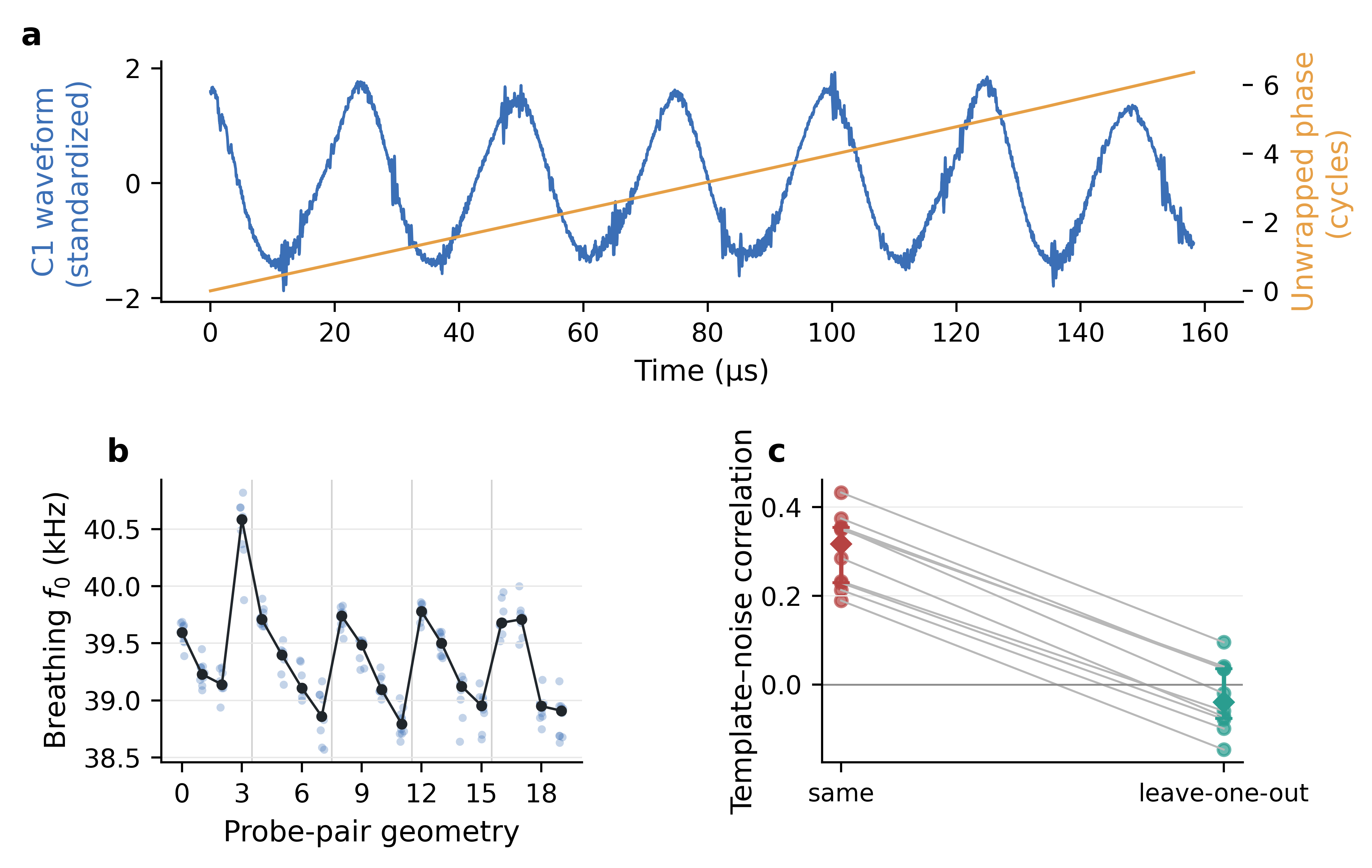}
  \caption{Breathing-phase coordinate and cross-fitted decomposition.
  (a) Representative correctly decoded fixed-probe waveform (C1) and its
  conditioning phase. (b) Record-level breathing-frequency estimates across
  the 20 geometries; points denote records and the line connects
  within-geometry medians. (c) Same-record and
  leave-one-record-out template--noise correlations for 10 synthetic records
  in one prespecified validation realization. Lines pair fits for each record;
  diamonds and bars show the median and interquartile range.}
  \label{si:fig:phase-crossfit}
  \alttext{Three panels connect the measured waveform to its breathing-phase coordinate, show record-level breathing frequencies across 20 geometries, and compare template--noise correlations in a synthetic validation. Leaving each record out reduces the correlation introduced by same-record fitting.}
\end{figure}

In a 10-record synthetic leakage test (Fig.~\ref{si:fig:phase-crossfit}(c)),
the mean correlation between each
record's phase-binned noise and the template error was 0.301 for same-record
fitting and $-0.027$ for leave-one-record-out fitting. The cross-fitted
third-order expansion had a normalized numerical closure error of
$1.7\times10^{-16}$. These tests validate the implemented data split and
algebra, not a physical model for $R$.

% Let S3 text fill the page while Figure S2 floats nearby.

% ------------------------------------------------------------
\section{Power accounting and phase-resolved residual-power robustness}
\label{si:sec:power-gating}
% ------------------------------------------------------------

\subsection{Spectral identity and two ratio conventions}

Equation~\eqref{si:eq:crossfit-decomposition} gives the exact spectral identity
\begin{equation}
  P_X(f)=P_C(f)+P_R(f)+P_{\mathrm{cross}}(f),
  \qquad
  P_{\mathrm{cross}}(f)=2\,\operatorname{Re}S_{CR}(f).
  \label{si:eq:power-closure-si}
\end{equation}
The five-band analysis used ratios of band-integrated analytic-signal powers.
Within each record, the three band-integrated terms were divided by the
measured-signal power. Equal-record averages were then formed within each
geometry, followed by separate medians across the 20 geometry summaries.
A separate 50--80~kHz check formed the three ratios at each frequency bin
before averaging over frequency. This per-bin convention uses different
weights and gave geometry-level medians of 0.75637, 0.23511, and 0.00940 for
$P_C/P_X$, $P_R/P_X$, and $P_{\mathrm{cross}}/P_X$, respectively.

The numerical closure audit was performed before geometry aggregation. For
each record $r$ and band $B$, the dimensionless phase-resolved closure error
was
\begin{equation}
 \epsilon_{rB}=\frac{\displaystyle\max_{\phi}\left|
 P_{X,rB}(\phi)-P_{C,rB}(\phi)-P_{R,rB}(\phi)
 -P_{\mathrm{cross},rB}(\phi)\right|}
 {\displaystyle\left\langle P_{X,rB}(\phi)\right\rangle_{\phi}}.
 \label{si:eq:phase-power-closure-si}
\end{equation}
The maximum over all 985 record--band combinations was
$\FreezePhasePowerClosureMax$. Because this audit preceded aggregation, it does
not require separately reported geometry medians to close exactly.

\subsection{Phase-resolved residual power and circular-shift test}

\begin{figure}[!htbp]
  \centering
  \includegraphics[width=\textwidth]{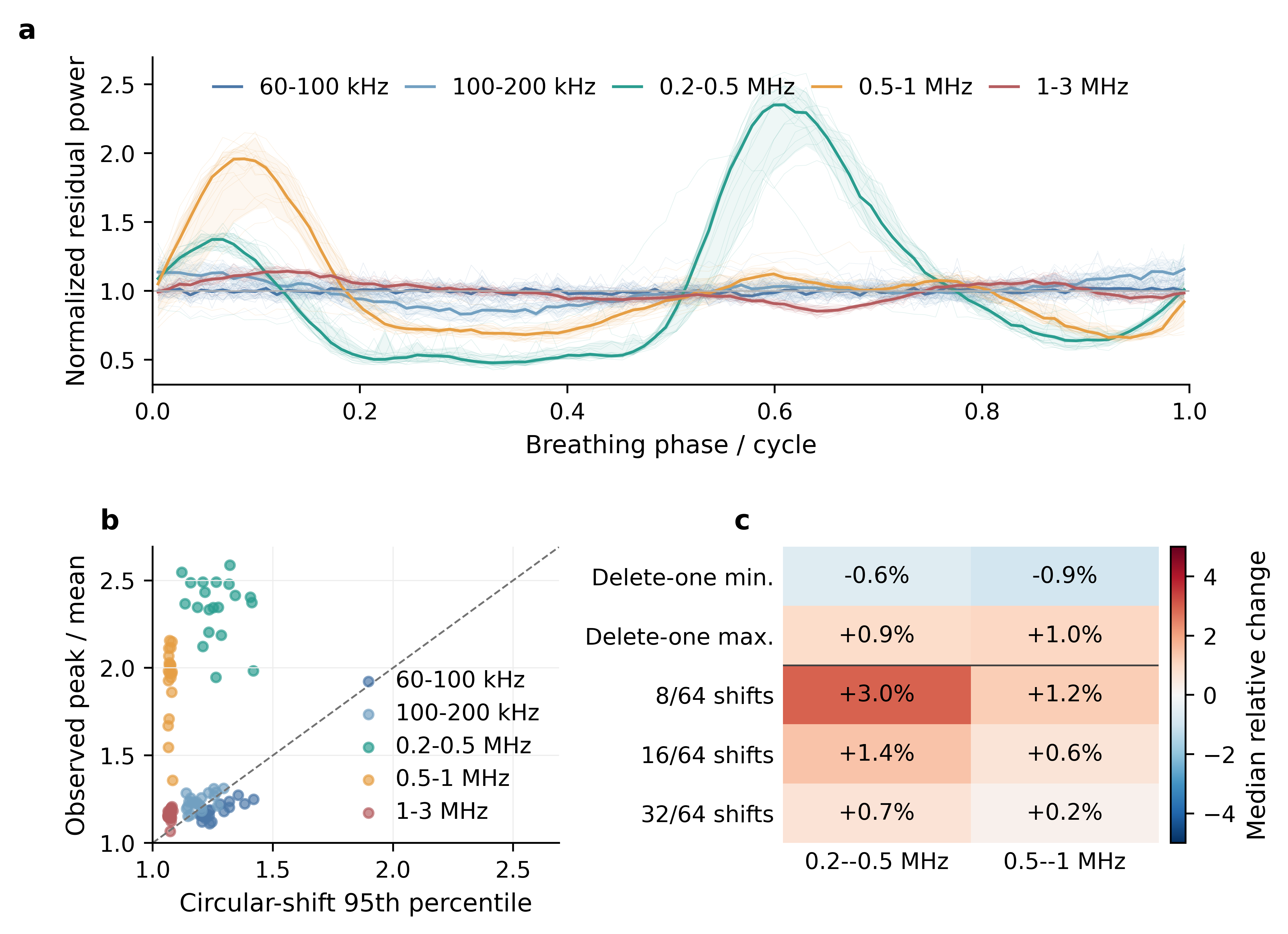}
  \caption{Residual-power organization and robustness. (a) Phase-resolved
  residual-power curves for all five bands and 20 geometries. Thin curves are
  individual geometries, bold curves are geometry medians, and shading spans
  the 10th--90th percentiles. (b) Observed peak-to-mean ratios against their
  geometry-specific 95th-percentile circular-shift thresholds; the diagonal marks
  equality. Main-text Table~I gives the corresponding exceedance counts.
  (c) Median relative changes in the two sub-MHz bands after record deletion
  and when 8, 16, or 32 shifts replace the 64-shift estimate of the mean
  shifted curve. The divider separates observed-statistic and circular-shift
  construction checks.}
  \label{si:fig:power-gating-robustness}
  \alttext{Phase-power curves and threshold comparisons summarize residual modulation across five frequency bands and 20 geometries. A third panel displays sensitivity to deleting records and changing the number of circular shifts used to estimate the mean shifted curve.}
\end{figure}

For frequency band $B$, the phase-resolved residual power and its peak-to-mean
ratio were defined as
\begin{align}
  P_{R,B}(\phi)
    &=\mathbb{E}\!\left[|z_{R,B}(t)|^2\mid\phi(t)=\phi\right],
    \label{si:eq:conditional-power-si}\\
  M_B
    &=\frac{\max_{\phi}P_{R,B}(\phi)}
            {\langle P_{R,B}(\phi)\rangle_{\phi}}.
    \label{si:eq:peak-mean-si}
\end{align}
Here $z_{R,B}$ is the band-limited complex envelope. This quantity is a
conditional second moment, not a conditional variance unless
$|\mathbb{E}[z_{R,B}\mid\phi]|^2$ vanishes. No such equality was assumed.

For each record and band, squared complex-envelope magnitudes were averaged
within 96 equal breathing-phase bins. Each curve was normalized by its
unweighted phase-bin mean and averaged with equal record weights within its
geometry. The statistic $M_B$ was calculated from this unsmoothed aggregate
curve, not from record-level peak-to-mean ratios.

Circular shifts paired unchanged residual-power samples at $t$ with
$\phi[(t+\tau)\bmod T]$, where $\tau$ is the circular time displacement and
$T$ is the record duration.
Each record used 64 independent displacements drawn uniformly between 5 and
95~ms, shared across its five bands. Records used independent draws, and each
shifted curve was renormalized by its own phase-bin mean.
For each band, 2000 null draws selected one shifted curve per record and
repeated geometry averaging and peak extraction. Their 95th percentile
defined the geometry-specific threshold. Separate 2000-draw record bootstraps
\cite{efron1979bootstrap} of unshifted curves gave intervals for the observed statistic.

The circular-shift test evaluates phase--power alignment conditional on the
existing decomposition.
Main-text Fig.~6(b) averages shifted curves for display, whereas Fig.~6(c)
shows median geometry-specific thresholds. Exceedance counts use each
geometry's threshold, not the displayed mean shifted curve or median threshold.

The robustness analysis deleted each complete record from the observed $M_B$
and varied the shifts used to estimate the mean shifted curve. The shift-count
check does not assess convergence of the 95th-percentile threshold or its
exceedance counts. The phase-extraction passband, 96-bin resolution, and five
power bands were held fixed. For the two sub-MHz bands, median delete-one-record
limits differed from the complete-record values by $-0.9\%$ to $+1.0\%$.
Reducing the circular-shift construction from 64 to eight shifts gave median
relative changes of up to 3.0\% in the peak of the mean shifted curve, with
smaller changes for 16 and 32 shifts
(Fig.~\ref{si:fig:power-gating-robustness}(c)).
Sensitivity to the 50--80~kHz signed-third-order target is reported in
Sec.~\ref{si:sec:third-order}.

\FloatBarrier

% ------------------------------------------------------------
\section{Bicoherence comparison under a common estimator}
\label{si:sec:bicoherence}
% ------------------------------------------------------------

\subsection{Estimator and prespecified harmonic region of interest}

For $U\in\{X,C,R\}$, auto-bicoherence was calculated using the same
estimator, segmentation, and frequency grid:
\begin{equation}
  b_U^2(f_1,f_2)=
  \frac{\left|\left\langle U(f_1)U(f_2)
  U^*(f_1+f_2)\right\rangle\right|^2}
  {\left\langle|U(f_1)U(f_2)|^2\right\rangle
   \left\langle|U(f_1+f_2)|^2\right\rangle}.
  \label{si:eq:bicoherence-si}
\end{equation}
Angle brackets denote averaging over non-overlapping, 10,000-sample
Hann-windowed segments. The signed input grid contained nonzero 10~kHz
multiples with $0<|f_1|,|f_2|\leq1.5$~MHz and
$0.05\leq f_1+f_2\leq1.5$~MHz. Negative-frequency coefficients were obtained
by complex conjugation. Each unordered pair was evaluated once and displayed
symmetrically; main-text Fig.~7(a--c) shows the central $\pm0.6$~MHz region.

\begin{figure}[H]
  \centering
  \includegraphics[width=\textwidth]{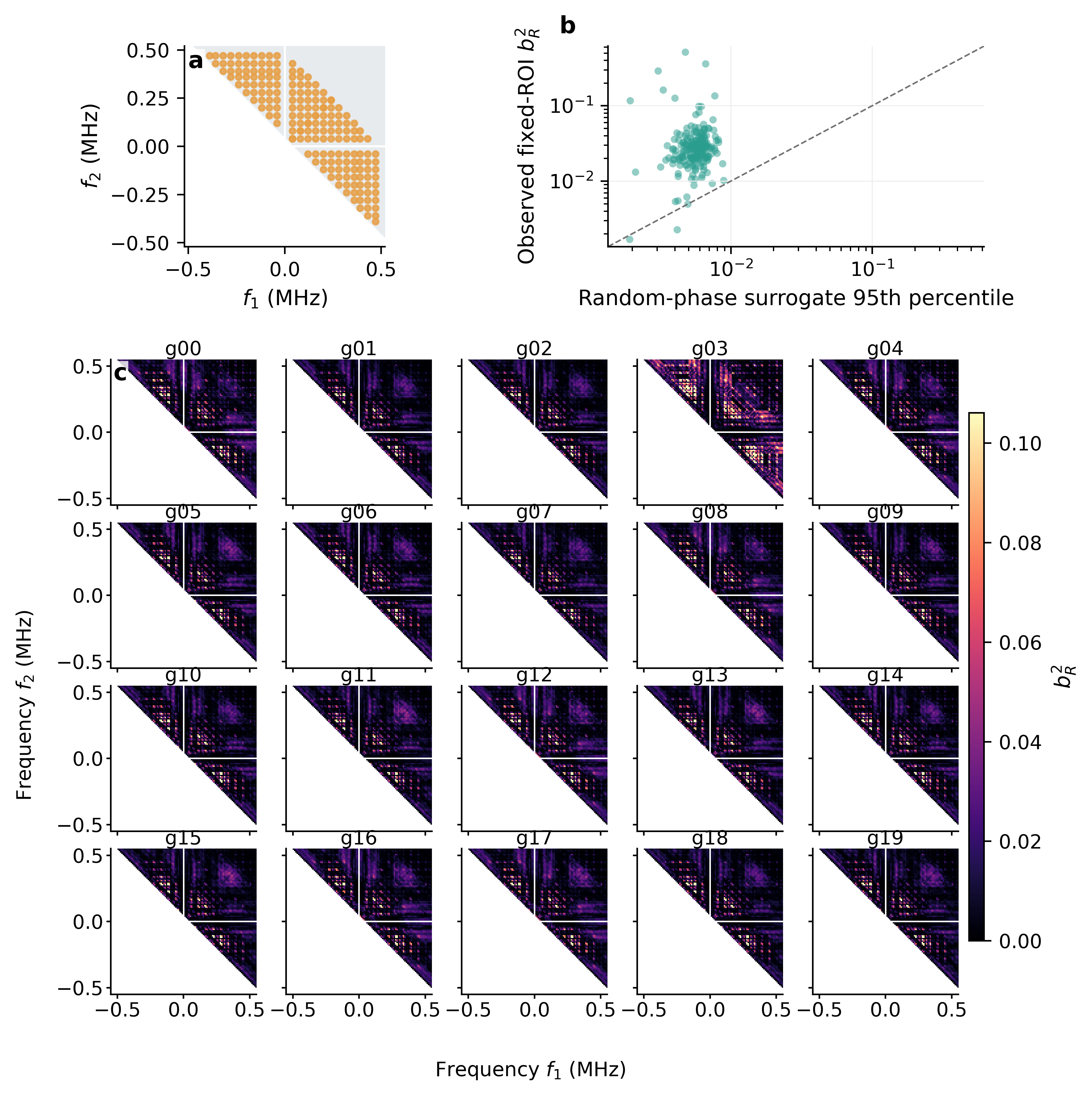}
  \caption{Complete residual-bicoherence audit. (a) Prespecified harmonic
  cells retained by the common $X$-power-support rule. ROI selection used no
  held-out bicoherence. (b) Record-level residual statistics against their
  record-specific 95th-percentile random-phase surrogate thresholds.
  (c) Residual maps for all 20 geometries with common axes and a
  shared within-$R$ scale. The maps compare residual morphology only; the
  absolute $X/C/R$ separation is reported in the text and main-text Fig.~7.}
  \label{si:fig:bicoherence-all}
  \alttext{The harmonic region of interest is displayed alongside record-level residual bicoherence and random-phase thresholds. A grid of 20 residual maps uses common axes and a shared residual-only color scale to compare structure across geometries.}
\end{figure}

The harmonic region of interest (ROI; Fig.~\ref{si:fig:bicoherence-all}(a)) was
fixed without using held-out bicoherence. It contained
integer sum- and difference-frequency triads whose positive target leg covered
harmonics 2--12 of the measured $f_0$.
Input harmonic orders $m,n$ satisfied $1\leq|m|,|n|\leq12$ and
$2\leq m+n\leq12$. Their frequencies were independently mapped to the
nearest grid bins, $a=\operatorname{round}(mf_0/\Delta f)$ and
$b=\operatorname{round}(nf_0/\Delta f)$, with $\Delta f=10$~kHz and
half-integer ties rounded to the nearest even integer.
The target bin was then set to $k=a+b$, rather than obtained by independently
rounding $(m+n)f_0/\Delta f$. Duplicate unordered pairs were removed, and
the frequency bounds stated above were applied.
All three legs were required to exceed
the common $X$-power threshold of $-60$~dB relative to the maximum over
0.05--1.5~MHz. The principal statistic was the $X$-denominator-weighted mean
$b^2$ over this ROI. An auxiliary coverage statistic counted the fraction of
$X$ target cells with $b_X^2>0.05$ covered by $C$ or $R$; this descriptive
threshold did not define the ROI or enter model calibration.

Each signal used its own normalization denominator in
Eq.~\eqref{si:eq:bicoherence-si}, whereas the ROI and $X$-derived scalar weights
were common. The comparison is therefore not an additive decomposition of
bicoherence. A value $b_C^2\simeq1$ indicates reproducible harmonic phase
relations, not that $C$ contains all signal energy.

\subsection{Random-phase surrogates and validation}

Each complete record used 200 residual surrogates that retained every
segment-by-frequency magnitude of $R$ and independently randomized Fourier
phase, following the Fourier-phase randomization principle of surrogate-data
testing \cite{theiler1992surrogate}. This preserved ROI-cell denominators while removing phase coherence.
The 95th percentile of the surrogate weighted-ROI statistic defined the
record-specific threshold. The test is conditional on the observed
spectral-amplitude sequence, not every correlated random process.

Validation returned $b_C^2\simeq1$ for a phase-locked harmonic chain,
$1.24\times10^{-4}$ for independent noise, and 0.9951 for an implanted
residual triad. All outputs satisfied $0\leq b^2\leq1$. The 20-geometry audit
confirmed that ROI selection did not use held-out bicoherence.

The ordering $C>X>R$ held for all 20 geometry summaries, and $C$ had greater
target-cell coverage than $R$ in every record. Figure~\ref{si:fig:bicoherence-all}
gives the complete residual-surrogate comparison using surrogate seed 20260902.
The residual statistic exceeded its record-specific threshold in
\FreezeResidualNullCount/\FreezeResidualNullTotal{} records.

\subsection{Fixed-ROI finite-sample and parameter sensitivity}

One-factor variants changed realization sampling (distributed halves or
quarters, or the first half), power-support floors ($-50$ or $-70$~dB), or
maximum harmonic order (8 or 10). Table~\ref{si:tab:bicoherence-sensitivity}
reports the corresponding record-level medians and geometry-cluster
bootstrap intervals, including the $1/N$-corrected residual statistic, where
$N$ is the number of spectral segments retained in the corresponding variant.
The tests did not vary fast Fourier transform (FFT) window length or overlap.

\begin{table}[H]
  \centering
  \renewcommand{\arraystretch}{1.25}
  \caption{Fixed-ROI bicoherence sensitivity to one-factor analysis variants.
  Entries are record-level medians, with 95\% intervals from resampling the
  20 geometry clusters in parentheses. Each variant contains 197 records;
  only the named setting differs from baseline.}
  \label{si:tab:bicoherence-sensitivity}
  \alttext{Rows compare the baseline bicoherence analysis with one-at-a-time setting changes. Columns report measured-signal, waveform, and residual summaries with geometry-cluster intervals, allowing the component ordering and sensitivity to be assessed.}
  \begin{tabularx}{\textwidth}{@{}>{\raggedright\arraybackslash}p{3.6cm}*{4}{>{\centering\arraybackslash}X}@{}}
    \toprule
    Analysis variant & $b_X^2$ & $b_C^2$ & $b_R^2$ & $b_R^2-1/N$ \\
    \midrule
    Baseline &
    \shortstack{0.9624\\{\scriptsize(0.9580--0.9661)}} &
    \shortstack{0.9991\\{\scriptsize(0.9989--0.9992)}} &
    \shortstack{0.0265\\{\scriptsize(0.0224--0.0292)}} &
    \shortstack{0.0255\\{\scriptsize(0.0216--0.0282)}} \\
    Distributed half-segments &
    \shortstack{0.9620\\{\scriptsize(0.9579--0.9660)}} &
    \shortstack{0.9991\\{\scriptsize(0.9989--0.9992)}} &
    \shortstack{0.0277\\{\scriptsize(0.0248--0.0315)}} &
    \shortstack{0.0257\\{\scriptsize(0.0230--0.0295)}} \\
    Distributed quarter-segments &
    \shortstack{0.9626\\{\scriptsize(0.9592--0.9665)}} &
    \shortstack{0.9991\\{\scriptsize(0.9989--0.9992)}} &
    \shortstack{0.0279\\{\scriptsize(0.0240--0.0320)}} &
    \shortstack{0.0239\\{\scriptsize(0.0200--0.0280)}} \\
    First half only &
    \shortstack{0.9624\\{\scriptsize(0.9582--0.9664)}} &
    \shortstack{0.9991\\{\scriptsize(0.9989--0.9992)}} &
    \shortstack{0.0261\\{\scriptsize(0.0236--0.0303)}} &
    \shortstack{0.0241\\{\scriptsize(0.0216--0.0283)}} \\
    Power floor $-50$~dB &
    \shortstack{0.9624\\{\scriptsize(0.9579--0.9661)}} &
    \shortstack{0.9991\\{\scriptsize(0.9989--0.9992)}} &
    \shortstack{0.0265\\{\scriptsize(0.0224--0.0292)}} &
    \shortstack{0.0255\\{\scriptsize(0.0214--0.0283)}} \\
    Power floor $-70$~dB &
    \shortstack{0.9624\\{\scriptsize(0.9579--0.9662)}} &
    \shortstack{0.9991\\{\scriptsize(0.9989--0.9992)}} &
    \shortstack{0.0265\\{\scriptsize(0.0224--0.0292)}} &
    \shortstack{0.0255\\{\scriptsize(0.0214--0.0282)}} \\
    Maximum harmonic order 8 &
    \shortstack{0.9624\\{\scriptsize(0.9579--0.9662)}} &
    \shortstack{0.9991\\{\scriptsize(0.9989--0.9992)}} &
    \shortstack{0.0265\\{\scriptsize(0.0224--0.0290)}} &
    \shortstack{0.0255\\{\scriptsize(0.0214--0.0282)}} \\
    Maximum harmonic order 10 &
    \shortstack{0.9624\\{\scriptsize(0.9579--0.9661)}} &
    \shortstack{0.9991\\{\scriptsize(0.9989--0.9992)}} &
    \shortstack{0.0265\\{\scriptsize(0.0224--0.0292)}} &
    \shortstack{0.0255\\{\scriptsize(0.0214--0.0282)}} \\
    \bottomrule
  \end{tabularx}
\end{table}

\subsection{Illustrative configuration and record selection}
\label{si:sec:illustrative-configuration}

The illustrative configuration was selected from the probe geometry before
examining the higher-order results. First, the median of the five nominal
scan angles, $45^{\circ}$, was selected. Among the four separations at this
angle (2.90, 3.43, 3.97, and 4.50~mm), 3.43~mm was closest to the
20-configuration median of approximately 3.65~mm. This configuration has
identifier 09 in the source data; Table~\ref{si:tab:probe-configurations}
maps all configuration identifiers to physical settings. The nominal scan
angles differ from the reconstructed separation directions because the stage
rotation center is offset from the fixed probe (Section~S1.2).

Main-text Fig.~4 uses record 73, the smallest complete record identifier in
this configuration, selected without reference to its spectral results.
The other nine records, identifiers 74--82, supplied its waveform template.
Main-text Fig.~7(a--d), Fig.~8(a), and the frequency-resolved spectrum in
Fig.~\ref{si:fig:third-order-robustness}(a) use the same configuration. The
main-text statistical summaries cover all 197 records or all 20 configuration
summaries, as specified in each figure caption.
Figure~\ref{si:fig:bicoherence-all} supplies the common ROI, all 20 residual maps,
and the record-level surrogate comparison.

\begin{table}[H]
  \centering
  \caption{Source-data identifiers and physical probe-pair settings. Angles
  are nominal scan commands; separations are reconstructed tip spacings,
  rounded to 0.01~mm. Identifiers label configurations, not acquisition order.}
  \label{si:tab:probe-configurations}
  \alttext{Twenty configurations combine five nominal scan angles and four translation settings. Identifier 09 corresponds to 45 degrees and a reconstructed tip separation of 3.43 millimeters.}
  \begin{tabular}{@{}rrr@{\qquad}rrr@{}}
    \toprule
    ID & Angle (deg) & Separation (mm) &
    ID & Angle (deg) & Separation (mm) \\
    \midrule
    00 & 0.0 & 3.00 & 10 & 45.0 & 3.97 \\
    01 & 0.0 & 3.53 & 11 & 45.0 & 4.50 \\
    02 & 0.0 & 4.07 & 12 & 67.5 & 2.81 \\
    03 & 0.0 & 4.60 & 13 & 67.5 & 3.34 \\
    04 & 22.5 & 2.97 & 14 & 67.5 & 3.87 \\
    05 & 22.5 & 3.50 & 15 & 67.5 & 4.41 \\
    06 & 22.5 & 4.04 & 16 & 90.0 & 2.70 \\
    07 & 22.5 & 4.57 & 17 & 90.0 & 3.23 \\
    08 & 45.0 & 2.90 & 18 & 90.0 & 3.77 \\
    09 & 45.0 & 3.43 & 19 & 90.0 & 4.30 \\
    \bottomrule
  \end{tabular}
\end{table}

\FloatBarrier

% ------------------------------------------------------------
\section{Exact signed third-order expansion and robustness}
\label{si:sec:third-order}
% ------------------------------------------------------------

\subsection{Trilinear identity and component groups}

To complement magnitude-based bicoherence, the signed analysis uses a
cross-fitted modified Ritz-type third-order functional
\cite{ritz1986,ritz1989,kim1996,brown2023prl,brown2023pre}. It combines
training-only regressor screening, leave-one-record-out fitting, a fixed target
band, and an exact $C/R$ expansion. Without a propagation-speed or
probe-separation factor, the normalized output is dimensionless rather than a
calibrated transfer rate or flux.

Let $\mathcal{G}_B(U,V,W)$ denote the cross-fitted signed third-order spectral
functional evaluated over band $B$. Within each held-out fold, the fitted
operator was fixed before evaluating any $C/R$ component. Trilinearity and
$X=C+R$ then give the exact identity
\begin{equation}
  \mathcal{G}_B(X,X,X)=
  \sum_{U,V,W\in\{C,R\}}\mathcal{G}_B(U,V,W).
  \label{si:eq:third-order-expansion-si}
\end{equation}
All eight terms were retained. The $CCC$ and $RRR$ groups contain only
waveform and residual inputs, respectively; one-$R$ and two-$R$ each sum
three permutations. The residual-associated contribution is the sum of
one-$R$, two-$R$, and $RRR$. These signed ratios are algebraic comparisons,
not energy fractions; the expansion is trilinear, not a cumulant decomposition.

For a held-out record $r$, the normalized functional used in the principal
comparison was
\begin{equation}
 \mathcal{G}_{B,r}(U,V,W)=
 \sum_{k:\,f_k\in B}\frac{1}{2P_{X,r}(k)}
 \operatorname{Re}\!\left[
 e^{-\mathrm{i}\Delta\theta_k^{(-r)}}
 \sum_{(a,b)\in\mathcal{Q}_k^{(-r)}}
 Q_{ab,k}^{(-r)}
 \left\langle U_aV_bW_k^*\right\rangle_r
 \right].
 \label{si:eq:signed-functional-si}
\end{equation}
For band $B$, $\mathcal{G}_{B,r}(X,X,X)$ is denoted by
$S_{\mathrm{total},r}$. Here $U,V,W\in\{C,R\}$ are fixed-probe components and
$P_{X,r}(k)=\langle|X_k|^2\rangle_r$. The coefficients $Q$ relate selected
fixed-probe quadratic products to the movable-probe response, and
$\Delta\theta_k^{(-r)}=\arg\langle Y_kX_k^*\rangle_{-r}$ is the training-set
cross-probe phase. Each fold estimated $Q$, $\Delta\theta$, and $\mathcal{Q}$
from the other complete records at that geometry. Signed bins obeyed $a+b=k$,
with $a,b\ne0$ and $|f_a|,|f_b|\leq1.5$~MHz. Division by $2P_X$ makes
Eq.~\eqref{si:eq:signed-functional-si} dimensionless.

Non-overlapping 10,000-sample Hann segments at 100~MS/s gave 10~kHz spacing.
The primary band contained \FreezeTargetBinCount{} bins at 50--80~kHz. Across
records, $2f_0$ ranged from \FreezeTwoFZeroRecordMinKHz{} to
\FreezeTwoFZeroRecordMaxKHz{}~kHz and crossed the upper edge for some records.
The statistic therefore samples the region near the second breathing harmonic
rather than tracking $2f_0$ in every record.

Candidate regressors required training-set bicoherence above
$\max[0.05,-\ln(0.01)/N_{\mathrm{train}}]$, where
$N_{\mathrm{train}}$ is the number of training spectral segments; at most 32
regressors were retained.
Before fitting, the linear projection of $Y_k$ on $X_k$ was removed, and each
quadratic regressor was projected off $X_k$. The primary and strict relative
singular-value cutoffs were $10^{-3}$ and $10^{-1}$. Templates and operators
excluded the held-out record. Because screening uses training-set
bicoherence, the attribution remains conditional on the fitted cross-validated
operator.

Conditional 95\% intervals used the 2.5th and 97.5th percentiles of 2000
record-bootstrap draws \cite{efron1979bootstrap}, keeping each record's component terms together.
Templates, spectra, fitted operators, cross-probe phases, and power
normalizations remained fixed.

\begin{figure}[!htbp]
  \centering
  \includegraphics[width=\textwidth]{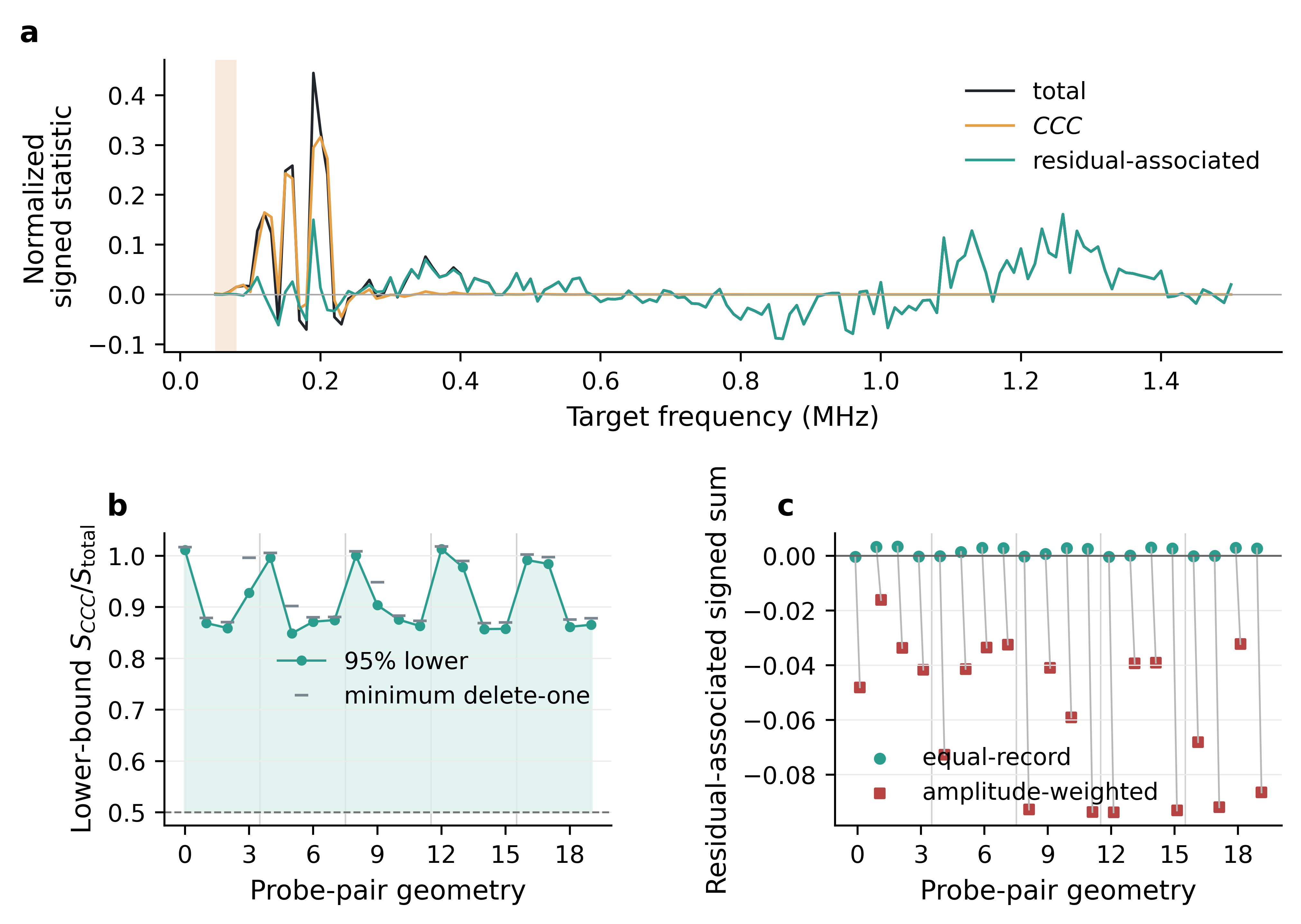}
  \caption{Signed third-order attribution and robustness. (a)
  Frequency-resolved total, $CCC$, and residual-associated sum through
  1.5~MHz for the illustrative configuration ($45^{\circ}$, 3.43~mm;
  Section~\ref{si:sec:illustrative-configuration}); shading marks the 50--80~kHz target.
  (b) Primary full-refit jackknife 95\% lower endpoints across 20 geometries;
  gray marks show minimum delete-one ratios and the dashed line marks 0.5.
  (c) Residual-associated sums under equal-record and
  record-amplitude weighting.}
  \label{si:fig:third-order-robustness}
  \alttext{Three panels show the broad signed third-order spectrum, geometry-level lower confidence endpoints for the all-waveform-to-total ratio, and residual-associated sums under two weighting conventions. The highlighted spectral interval identifies the fixed attribution target.}
\end{figure}

The separate full-refit jackknife \cite{miller1974jackknife} deleted each record and rebuilt all remaining
templates, spectra, regressors, $Q$, cross-probe phases, and normalized
contributions. Only record-level breathing frequencies and phases remained
fixed. For $n$ records, variance was $(n-1)/n$ times the summed squared
deviations of deletion ratios from their mean. Approximate two-sided 95\% Student-$t$
intervals used $n-1$ degrees of freedom and were centered on the full-sample
ratio. Reported lower bounds are the lower interval endpoints.
Figure~\ref{si:fig:third-order-robustness}(a,b) shows the frequency-resolved
contributions and geometry-level jackknife lower endpoints.
Additional checks varied aggregation, band placement, frequency resolution,
operator threshold, harmonic-set size, and envelope-preserving Gaussian
construction. Segments and cycles were not independent experimental replicates.

\subsection{Dominant phase-locked contribution and residual boundary}

Table~\ref{si:tab:third-order-summary} gives the exact primary and strict results.
The normalized total was positive, and $CCC/\mathrm{total}>0.5$, for all 20
geometries under both cutoffs.

The attribution persisted under target-band checks. Excluding the bin nearest
record-wise $2f_0$ gave a median/minimum ratio of
\FreezeExcludeTwoFCCCMedian/\FreezeExcludeTwoFCCCMinimum. Bands of 40--80 and
50--90~kHz gave \FreezeShiftFortyEightyCCCMedian/
\FreezeShiftFortyEightyCCCMinimum{} and
\FreezeShiftFiftyNinetyCCCMedian/\FreezeShiftFiftyNinetyCCCMinimum.
$CCC$ remained dominant in \FreezeHigherTwoThreeCount/20 geometries at
$2.25$--$3.0f_0$ and \FreezeHigherThreeFourCount/20 at $3.25$--$4.0f_0$.
These extensions do not establish attribution at all harmonics.

\begin{table}[H]
\centering
\caption{Geometry-level robustness of the phase-locked contribution. Ratios
are signed trilinear comparisons, not energy fractions.}
\label{si:tab:third-order-summary}
  \alttext{Geometry-level summaries compare all-waveform-to-total ratios under primary and strict truncations, their robustness checks, and residual-associated comparisons. The values describe signed statistical contributions and do not represent fractions of physical energy.}
\begin{tabular}{@{}lrr@{}}
\toprule
Quantity & Primary cutoff & Strict cutoff \\
\midrule
Median normalized total & \FreezeSignedPrimaryTotal{} & \FreezeSignedStrictTotal{} \\
Median $CCC/\mathrm{total}$ & \FreezeSignedPrimaryCCCFrac{} & \FreezeSignedStrictCCCFrac{} \\
Minimum $CCC/\mathrm{total}$ & \FreezeSignedPrimaryCCCFracMinimum{} & \FreezeSignedStrictCCCFracMinimum{} \\
Minimum full-refit jackknife lower bound & \FreezeJackknifePrimaryLowerMinimum{} & \FreezeJackknifeStrictLowerMinimum{} \\
Positive total & \FreezeSignedPrimaryPositiveCount/20 & \FreezeSignedStrictPositiveCount/20 \\
$CCC/\mathrm{total}>0.5$ & \FreezeSignedPrimaryCCCDominanceCount/20 & \FreezeSignedStrictCCCDominanceCount/20 \\
\bottomrule
\end{tabular}
\end{table}

For the primary cutoff, median normalized one-$R$, two-$R$, and $RRR$ terms
were $+\FreezeSignedPrimaryOneR$, $+\FreezeSignedPrimaryTwoR$, and
$\FreezeSignedPrimaryRRR$. The two-$R$ and $RRR$ signs were uniform, but their
residual-associated sum was positive in
\FreezeSignedPrimaryResidualPositiveCount/20 geometries. Under
record-amplitude weighting, the unnormalized bootstrap interval was negative
in 18/20 geometries and crossed zero in two. The aggregate sign therefore
depends on weighting and normalization
(Fig.~\ref{si:fig:third-order-robustness}(c)).

The envelope-preserving Gaussian null matched each record's paired residual
auto- and cross-spectra and channel-specific second moments in 96 phase bins.
Phase trajectories used measured segment-start phases and record-level $f_0$.
Paired complex-Gaussian coefficients were spectrally matched, then subjected
to two alternating rounds of phase-bin amplitude rescaling and spectral
rematching. Without a convergence criterion, envelope preservation denotes
approximate second-moment matching, not pointwise preservation.

For each geometry, 200 surrogate data sets retained $C_X/C_Y$ and replaced
only the paired residuals. Each surrogate repeated leave-one-record-out
regressor selection and refitting of $Q$ and cross-probe phase from generated
$C+R$ signals. The statistic was the equal-record mean of
$S_{1R}+S_{2R}+S_{RRR}$ with the frequency-wise normalization in
Eq.~\eqref{si:eq:signed-functional-si}. No division by $S_{\mathrm{total}}$,
absolute value, square, or norm was applied.
Its surrogate 95th percentile defined the geometry-specific threshold.
Constraints used each record's measured residuals, so this was not a
training-only prediction test.

The final spectral rematch can perturb phase-bin moments. Diagnostics were
retained only for the first surrogate at each geometry, not all 200.
The observed statistic exceeded the threshold in
\FreezeEnvelopeNullCount/\FreezeEnvelopeNullTotal{} geometries.
This specified Gaussian null did not reproduce all residual third-order structure;
the test does not identify its physical mechanism.

\FloatBarrier

% ------------------------------------------------------------
\section{Conditional comparison with Gaussian controls}
\label{si:sec:model}
% ------------------------------------------------------------

This supplementary comparison asks whether retaining the waveform and
constraining Gaussian residuals by measured spectral covariance and phase
gates reproduces the evaluated higher-order statistics. Constraint fidelity
is assessed before the generated outputs are interpreted. The comparison is
conditional on the existing decomposition and supplied breathing phase;
failure of a specified construction does not identify a plasma mechanism.

\subsection{Stationary and phase-modulated Gaussian controls}
\label{si:sec:model-construction}

Both controls retained the cross-fitted waveform and replaced the residual by
a zero-mean two-probe Gaussian process. M0 preserved residual spectral
covariance without phase modulation. M1 also applied fixed-probe
residual-power gates in 60--100~kHz, 100--200~kHz, 0.2--0.5~MHz, 0.5--1~MHz,
and 1--3~MHz. In compact notation,
\begin{equation}
  Z(t)=C_Z[\phi_X(t)]+\sum_b A_b[\phi_X(t)]\eta_{Z,b}(t),
  \qquad Z\in\{X,Y\},
  \label{si:eq:model-si}
\end{equation}
where $\eta_{Z,b}$ is constrained by the fold-specific covariance below. M0
uses constant $A_b$, whereas M1 uses the measured gates. ``Stationary'' refers
to the M0 residual, not the complete signal containing $C_Z$. Neither control
is a dynamical plasma model.

M1 used \FreezeMOneRealizations{} realizations and M0 used
\FreezeMZeroRealizations, with seed 20260902. Record $r$ supplied its segment
start phases, $f_0$, and valid segment count. Its template excluded $r$, and
its direct covariance and gate weights were zero. However, precomputed
residuals from another record $j$ used $C_{Z,g,-j}$, which can include $r$.
These upstream templates were not rebuilt inside the generation fold. The
reference distributions are therefore conditional on the existing
cross-fitted decomposition, not fully nested out-of-sample predictions.

The residual covariance used the same non-overlapping, 10,000-sample Hann
spectra. Independent unit-exponential draws were normalized over the training
records to generate Dirichlet$(1,\ldots,1)$ weights, corresponding to the
Bayesian bootstrap \cite{rubin1981bayesian}. At $f_k$, the
target was
\begin{equation}
 \bm S_k^{\mathrm{train}}=
 \begin{bmatrix}
 S_{XX,k} & S_{XY,k}\\
 S_{XY,k}^{*} & S_{YY,k}
 \end{bmatrix}.
 \label{si:eq:m1-covariance}
\end{equation}
Here, $S_{XX,k}=\langle R_{X,k}R_{X,k}^{*}\rangle$,
$S_{YY,k}=\langle R_{Y,k}R_{Y,k}^{*}\rangle$, and
$S_{XY,k}=\langle R_{X,k}R_{Y,k}^{*}\rangle$. Record-level segment averages
were combined with the bootstrap weights. The diagonal was real and
non-negative, and the off-diagonal entries were a conjugate pair. If
$|S_{XY,k}|>\sqrt{S_{XX,k}S_{YY,k}}$, only its magnitude was clipped to the
positive-semidefinite boundary. Paired complex-normal innovations were mapped
through this covariance without additional shrinkage.

The fixed-probe power gates were mean-normalized, square-rooted into amplitude
gates, and normalized to unit root-mean-square. The same gate was applied to
both channels. Synthesis used disjoint masks of 0--100, 100--200, 200--500,
500--1000, and $\geq1000$~kHz without transition bands. The retained support
ended at 1.51~MHz, including one 10~kHz padding bin; evaluation excluded the
padding bin and ended at 1.5~MHz. The first gate was extrapolated below
60~kHz. The fifth gate was calibrated over 1--3~MHz but evaluated only over
1--1.5~MHz.

Each gate used 96 phase bins, Fourier orders 0--12, a floor of 2\% of mean
amplitude, and root-mean-square normalization. The held-out record's start
phases and $f_0$ set the phase trajectory within each 0.1~ms segment. After
gating, one algebraic frequency-wise rematch restored the target sample
covariance. This differs from the two-round envelope-preserving Gaussian null
in Section~S5 and
does not define a frequency-resolved conditional covariance model.

Band, fixed-ROI bicoherence, and signed-third-order statistics were recomputed
from generated $C+R$ signals, rather than replaced by calibration targets.
Each data set refitted $Q$, cross-probe phase, regressor selection, and $P_X$
normalization. Generated $CCC$ non-coverage therefore diagnoses operator-refit
sensitivity, not loss of the retained waveform.

\begin{figure}[!htbp]
  \centering
  \includegraphics[width=\textwidth]{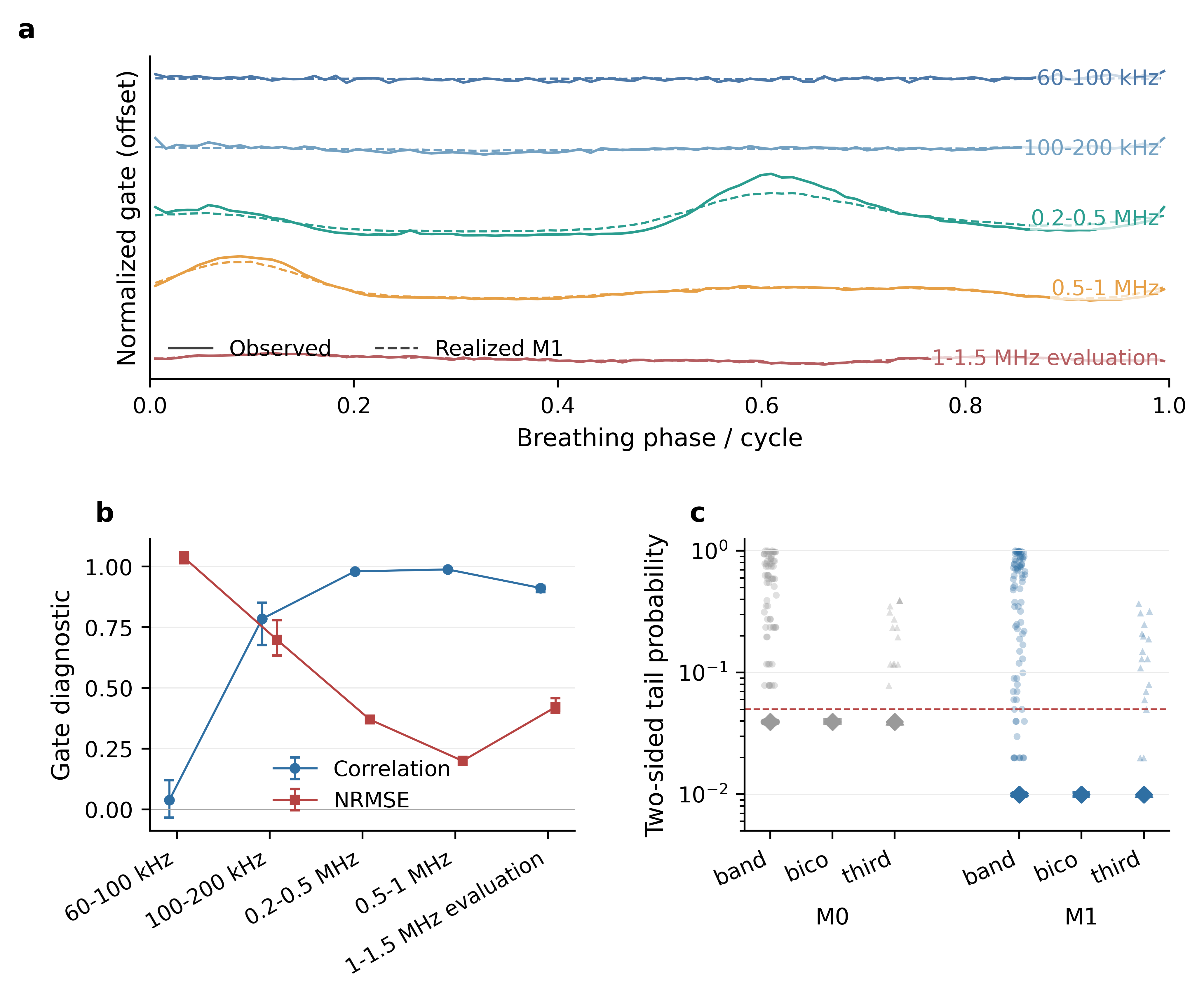}
  \caption{Gaussian-control diagnostics. (a) Target and realized M1 power
  gates in the representative geometry; solid and dashed curves denote the
  observed and realized profiles, respectively. (b) Gate-correlation and
  normalized root-mean-square error (NRMSE) medians with interquartile ranges
  across 197 records. The fifth output
  is evaluated over 1--1.5~MHz using a gate calibrated over 1--3~MHz.
  (c) Geometry-level two-sided tail probabilities for band, bicoherence, and
  signed third-order statistics under M0 and M1; diamonds mark medians and the
  dashed line marks 0.05. M0 and M1 used
  \FreezeMZeroRealizations{} and \FreezeMOneRealizations{} realizations.}
  \label{si:fig:model-diagnostics}
  \alttext{Observed and generated phase-power profiles, gate-fit diagnostics, and conditional tail probabilities assess the Gaussian controls. Agreement varies by frequency band; the lowest-frequency gate is poorly reproduced, limiting interpretation of the higher-order comparisons.}
\end{figure}

\subsection{Numerical constraints and gate reproduction}

Implementation checks covered covariance, bicoherence, direct-weight
exclusion, and seed reproducibility. The direct-weight check did not test the
indirect template dependence described above. For generated covariance
$\widehat S$ and training target $S$, the post-rematch error was
\begin{equation}
 \epsilon_2=\frac{\|\widehat S-S\|_2}{\|S\|_2},
 \label{si:eq:m1-l2-error}
\end{equation}
The norm covers non-negative bins through 1.5~MHz. A fixed diagnostic draw
with equal training weights was evaluated once per held-out record, separately
from the generated-output ensemble. Across 197 records, median errors for
$S_{XX}$, $S_{YY}$, and $S_{XY}$ were $\FreezeMOnePsdXMedian$,
$\FreezeMOnePsdYMedian$, and $\FreezeMOneCrossMedian$; the maximum was
$\FreezeMOneCovarianceMaximum$.

For each held-out record and band $B$, $p_B$ and $\widehat p_B$ are the target
and realized fixed-probe phase-power profiles, each normalized to unit phase
mean.
With equal weights over 96 phase bins, the normalized root-mean-square error
was
\begin{equation}
 \mathrm{NRMSE}_B=
 \frac{\sqrt{\langle(\widehat p_B-p_B)^2\rangle_\phi}}
 {\max\{\sigma_\phi[p_B],10^{-300}\}},
 \qquad
 \sigma_\phi[p_B]=
 \sqrt{\langle(p_B-\langle p_B\rangle_\phi)^2\rangle_\phi}.
 \label{si:eq:gate-nrmse-si}
\end{equation}
Here, $\langle\cdot\rangle_\phi$ denotes the 96-bin mean and $\sigma_\phi$ is
the population standard deviation. NRMSE is therefore relative to the target
phase modulation. Medians and interquartile ranges pool all 197 diagnostics.

Figure~\ref{si:fig:model-diagnostics}(b) summarizes gate correlations and NRMSE.
The record-median gate correlation was 0.038 in the 60--100~kHz band,
compared with 0.980 and 0.988 in the 0.2--0.5 and 0.5--1~MHz bands.
These correlations describe profile shape, not agreement in modulation depth.
Under the generated-output evaluation, observed peak-to-mean ratios fell
within M1's 95\% intervals in 0/20 and 4/20 geometries in the two sub-MHz
bands, respectively. The 60--100~kHz mismatch also limits interpretation of
the 50--80~kHz third-order target, which partly uses gate extrapolation below
60~kHz. Observed peak/mean medians were
\FreezeMOnePeakObservedByBand, compared with
\FreezeMOnePeakGeneratedByBand{} for M1. M0 gave
\FreezeMZeroSubMHzPeakGenerated{} in the two sub-MHz bands. These metrics were
recomputed from generated $C+R$ segments on the 0--1.5~MHz evaluation support;
they are distinct from the native-rate circular-shift estimator in Section~S3.

\subsection{Conditional generated-output assessment}

An observation was covered when it fell within the generated 2.5th--97.5th
percentile interval for its geometry and metric. Coverage is descriptive and
conditional on the existing decomposition and supplied phase. Primary and
strict third-order results were kept separate. Unequal realization counts
preclude formal aggregate comparison of M0 and M1.

\begin{figure}[!htbp]
  \centering
  \includegraphics[width=\textwidth]{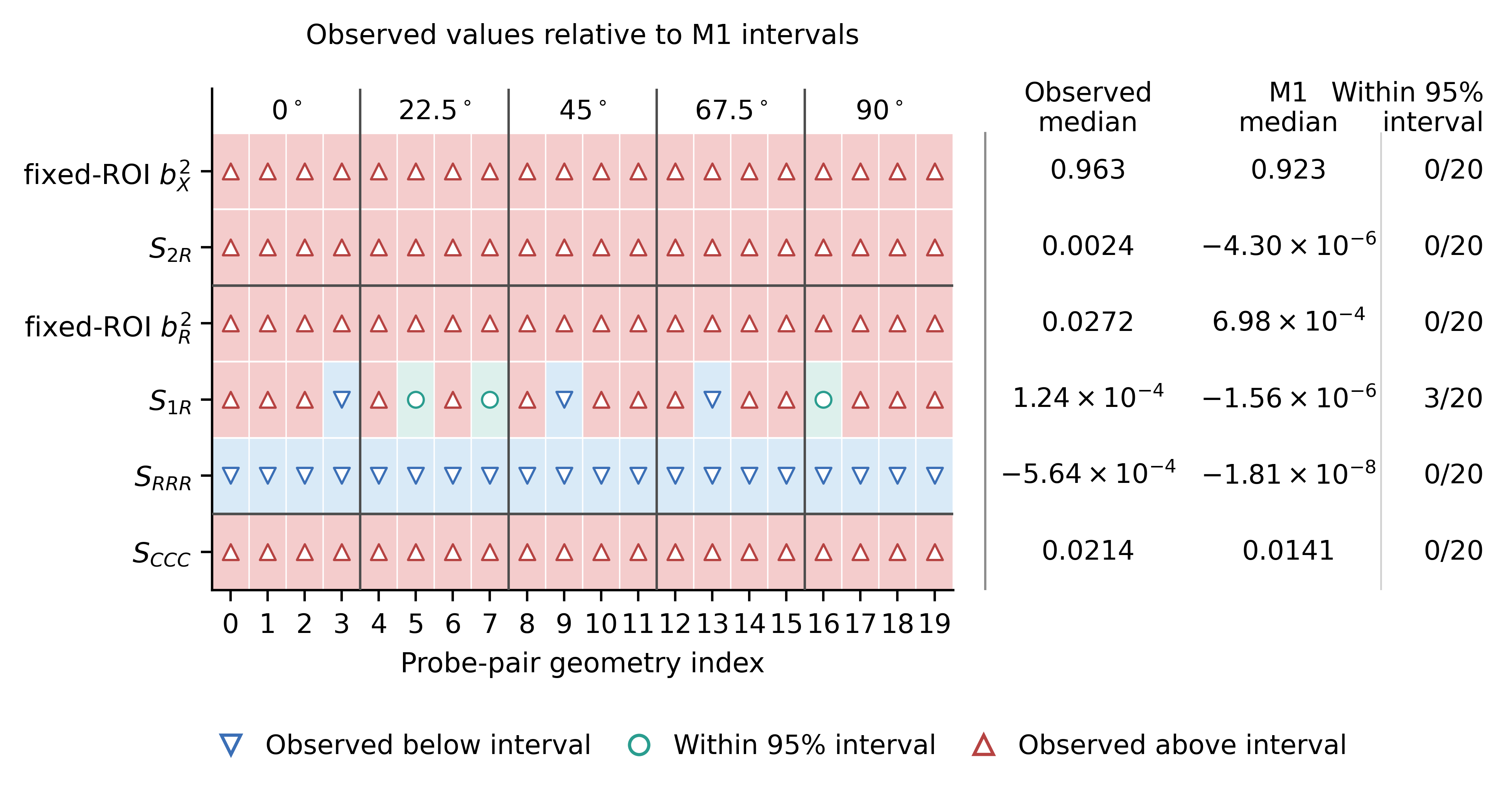}
  \caption{Conditional higher-order comparison with the phase-modulated
  Gaussian control (M1). Positions of the observations relative to M1
  simulation-based 95\% intervals across 20 geometries. The rows distinguish
  outputs not constrained to zero ($b_X^2$ and $S_{2R}$), outputs associated
  with odd residual moments ($b_R^2$, $S_{1R}$, and $S_{RRR}$), and
  refit-sensitive $S_{CCC}$. Symbols indicate whether an observed value lies
  below, within, or above its interval. Adjacent columns report geometry
  medians of observed and M1-generated outputs, followed by counts within
  the intervals. The signed third-order rows use the primary truncation;
  M1 uses \FreezeMOneRealizations{} realizations. The intervals are conditional
  on the existing decomposition and its indirect record dependence
  (Section~\ref{si:sec:model-construction}).}
  \label{si:fig:model-higher-order-check}
  \alttext{A geometry-by-statistic symbol matrix shows whether observations lie below, within, or above M1 intervals. Adjacent numerical columns give observed and generated medians and coverage counts, with rows grouped by their residual-moment and refitting roles.}
\end{figure}

Figure~\ref{si:fig:model-higher-order-check} reports the effect sizes, deviation
directions, and coverage counts for M1. The three groups distinguish outputs
not constrained to zero by residual symmetry, outputs associated with odd
residual moments, and the operator-refit diagnostic.

\begin{table}[!htbp]
\centering
\caption{Conditional geometry-level coverage by M0 and M1. The nonzero group
contains $b_X^2$ and two-$R$; the odd-moment group contains $b_R^2$, one-$R$,
and $RRR$, each under both truncations. $CCC$ diagnoses operator refitting.}
\label{si:tab:model-coverage}
  \alttext{Coverage counts compare M0 and M1 across band-power, bicoherence, and signed third-order outputs. Rows separate nonzero, odd-residual-moment, and operator-refitting comparisons; the intervals remain conditional on the existing decomposition.}
\begin{tabularx}{\textwidth}{@{}>{\raggedright\arraybackslash}p{3.9cm}>{\raggedright\arraybackslash}p{1.7cm}>{\raggedright\arraybackslash}p{1.7cm}>{\raggedright\arraybackslash}X@{}}
\toprule
Metric family & M0 coverage & M1 coverage & M1 non-coverage pattern \\
\midrule
Five-band fractions and peak/mean ratios & \FreezeMZeroBandCoverage{} (19.7\%) &
\FreezeMOneBandCoverage{} (22.0\%) & The two sub-MHz peak/mean outputs had M1
coverage of 0/20 and 4/20. \\
Higher-order outputs not constrained to zero & \FreezeMZeroNotZeroCoverage{} (0\%) & \FreezeMOneNotZeroCoverage{} (0\%) &
Primary $b_X^2$ and two-$R$ were above the M1 interval in 20/20 geometries
for each output. \\
Odd-residual-moment-constrained outputs & \FreezeMZeroZeroOddCoverage{} (13.0\%) & \FreezeMOneZeroOddCoverage{} (13.0\%) &
Primary $b_R^2$, one-$R$, and $RRR$ coverage was 0/20, 3/20, and 0/20. \\
$CCC$ signed-operator refit diagnostic & \FreezeMZeroCCCCoverage{} (0\%) & \FreezeMOneCCCCoverage{} (0\%) &
The generated operator and $P_X$ normalization were re-estimated, so this
row is not a direct test of whether $C$ was preserved. \\
\bottomrule
\end{tabularx}
\end{table}

For $b_X^2$, the observed and M1 medians were
\FreezeBicoXObserved{} and \FreezeBicoXMOne. For primary two-$R$, the medians
were \FreezeTwoRObserved{} and $\FreezeTwoRMOne$. Table~\ref{si:tab:model-coverage} reports the geometry-level non-coverage.

A centered Gaussian residual has zero population odd moments, although
finite-sample $b_R^2$ remains positive. Observed/M1 medians were
\FreezeBicoRObserved/\FreezeBicoRMOne{} for $b_R^2$,
$\FreezeOneRObserved$/$\FreezeOneRMOne$ for one-$R$, and
$\FreezeRRRObserved$/$\FreezeRRRMOne$ for $RRR$. Coverage was 2/197 records,
3/20 geometries, and 0/20 geometries, respectively. The specified control did
not reproduce these residual higher-order outputs.
The residual-bicoherence medians differ by approximately
\FreezeBicoRObservedOverMOne-fold under this common generated-output
evaluation; this contrast alone does not identify the source of the discrepancy.

The observed/M1 $CCC$ median was \FreezeCCCObserved/\FreezeCCCMOne;
its operator-refit interpretation is specified in Section~\ref{si:sec:model-construction}.

\subsection{Interpretation limits}

The comparison remains limited by indirect record dependence, shared
fixed-probe gates, low-frequency gate mismatch, and the 1--1.5~MHz evaluation
of a gate calibrated over 1--3~MHz. Channel-specific gates,
phase-conditioned cross-spectra, cross-band structure, and non-Gaussian
innovations were not tested.
Possible explanations therefore include finer phase-conditioned second-order
structure, residual distribution or cross-frequency structure, and effects of
operator refitting. The present comparison does not distinguish these
possibilities or determine whether any one extension would remove the
discrepancies. It establishes the limitation of the specified constructions,
not the insufficiency of every second-order model or a unique physical cause.

\FloatBarrier

% ------------------------------------------------------------
\section{Phase-reference and acquisition-order controls}
\label{si:sec:reviewer-controls}
% ------------------------------------------------------------

\subsection{Reciprocal phase-reference analysis}

The leave-one-record-out analysis was repeated for all four phase--target
combinations, $\phi_X\!\rightarrow X$, $\phi_X\!\rightarrow Y$,
$\phi_Y\!\rightarrow X$, and $\phi_Y\!\rightarrow Y$.
Peak-phase comparisons circularly aligned $\phi_Y$ to $\phi_X$ using each
record's mean segment-start phase difference.
The harmonic ROI used only $f_0$ and target-signal power support, not
observed bicoherence.

\begin{figure}[!htbp]
  \centering
  \includegraphics[width=\textwidth]{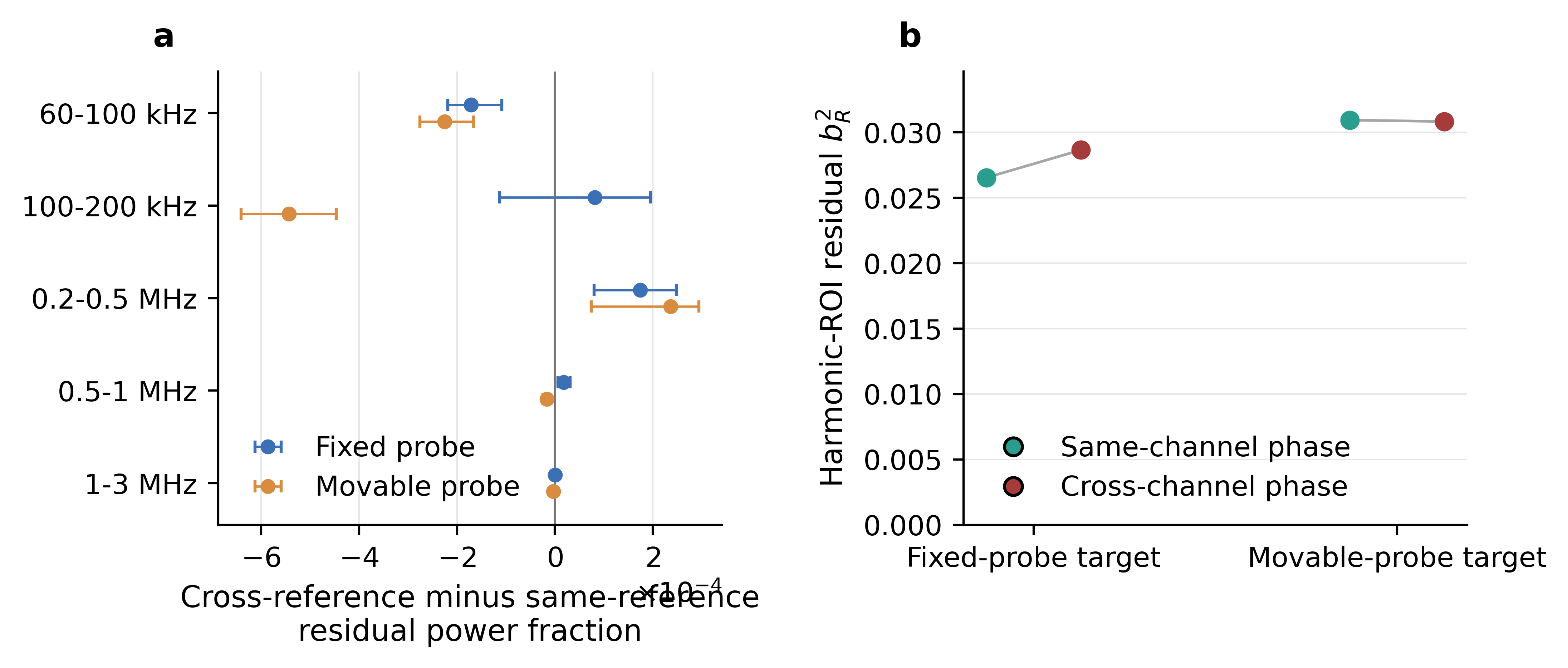}
  \caption{Reciprocal phase-reference control for 197 records in 20 geometries.
  (a) Median paired residual-power-fraction differences (cross-reference minus
  same-reference); bars show geometry-cluster bootstrap 95\% intervals.
  (b) Record-level median residual bicoherence under both references in the
  fixed harmonic ROI.}
  \label{si:fig:cross-channel-phase}
  \alttext{Two panels compare residual statistics when each probe uses its own or the other probe's breathing phase. Paired power-fraction differences include geometry-cluster intervals; bicoherence values show the effect of exchanging the conditioning reference.}
\end{figure}

Point estimates summarize records directly. Paired differences are medians
of record-wise cross-reference minus same-reference differences, not
differences between marginal medians.
Bootstrap draws resampled the 20 geometry clusters with replacement, retaining
all paired records within each selected cluster. Percentile intervals describe
the median of the pooled paired differences.
Figure~\ref{si:fig:cross-channel-phase} compares residual power fractions and
bicoherence under the same- and cross-channel phase references.

In the 0.2--0.5 and 0.5--1~MHz bands, fixed-probe residual peak-to-mean
medians were 2.4635/2.0573 (same-channel reference) and 2.4584/2.0713
(cross-channel reference). Movable-probe values were 2.2673/2.2553 and
2.2501/2.2689, respectively. All four paired differences had geometry-cluster
bootstrap 95\% intervals containing zero.
Median absolute peak-phase shifts after alignment were 0.0055--0.0057 cycles.
Paired median changes in sub-MHz coherent and residual power fractions
were below $6.0\times10^{-4}$ in magnitude.

Residual bicoherence medians under same- and cross-channel references were
0.02654/0.02866 for the fixed probe and 0.03094/0.03083 for the movable probe.
Paired median differences were 0.00280 (95\% interval 0.00079--0.00402)
and $-0.00098$ ($-0.00205$--0.00098), respectively.
The fixed-probe difference indicates local phase-reference sensitivity;
the scale separation from $X/C$ bicoherence persisted.

\FloatBarrier

\subsection{Acquisition-order sensitivity}

Without absolute timestamps, the file identifier served as a within-geometry
order proxy.
Order was scaled to $[-0.5,0.5]$ within each geometry, and slopes were
estimated after removing geometry-specific intercepts. Geometry-cluster
bootstraps retained the nesting of records.
All eight 95\% intervals included zero (Table~\ref{si:tab:acquisition-order}),
indicating no consistent within-geometry order trend.
This proxy cannot resolve time-dependent drift or turn sequential geometries
into simultaneous spatial measurements.

\begin{table}[H]
  \centering
  \caption{Within-geometry acquisition-order analysis. Slopes are per scaled
  geometry-order span after removal of geometry-specific intercepts. The
  $f_0$ slope is in kHz per span; all other slopes are dimensionless per span.
  Intervals use geometry-cluster bootstrap resampling.}
  \label{si:tab:acquisition-order}
  \alttext{Eight within-geometry slopes summarize changes in breathing frequency, band-power fractions, residual bicoherence, and the all-waveform-to-total ratio with acquisition order. Every geometry-cluster confidence interval includes zero, leaving no detected common linear trend.}
  \begin{tabularx}{\textwidth}{@{}>{\raggedright\arraybackslash}Xrrc@{}}
    \toprule
    Outcome & Slope & 95\% interval & \shortstack{Interval excludes\\zero} \\
    \midrule
    \SIAcquisitionOrderRows
    \bottomrule
  \end{tabularx}
\end{table}
\FloatBarrier

% ------------------------------------------------------------
\section*{Data provenance of supplementary figures}
% ------------------------------------------------------------

The experimental supplementary figures use the same 197 paired records
across 20 probe configurations as the main-text analyses. The source-data
index links all 394 channel files to their record identifiers, acquisition
headers, and SHA-256 hashes; numerical products retain processing settings.
The decoding check in Fig.~\ref{si:fig:encoding-audit} identified its inputs by
file paths and headers without per-file hashes. The example trace in
Fig.~\ref{si:fig:phase-crossfit} was independently matched to the source-data
index by its file hash.

\end{document}